\documentclass[aps,prd,nofootinbib,twocolumn,superscriptaddress,preprintnumbers,balancelastpage,longbibliography]{revtex4-1}

\usepackage[utf8]{inputenc}
\usepackage{amsmath}
\usepackage{amsfonts} 
\usepackage{amssymb}
\usepackage{graphicx}
\usepackage[dvipsnames]{xcolor}
\usepackage{hyperref}
\usepackage{tensor}
\usepackage{siunitx}
\usepackage{physics}
\usepackage{appendix}
\usepackage{cancel}

\hypersetup{colorlinks=true,
			urlcolor=purple,
			citecolor=blue,
			linkcolor=magenta,}

\newcommand{\beq}{\begin{equation}}
\newcommand{\eeq}{\end{equation}}
\newcommand{\be}{\begin{eqnarray}}
\newcommand{\ee}{\end{eqnarray}}
\newcommand{\bea}{\begin{eqnarray}}
\newcommand{\eea}{\end{eqnarray}}
\newcommand{\bi}{\begin{itemize}}
\newcommand{\ei}{\end{itemize}}
\newcommand{\ben}{\begin{enumerate}}
\newcommand{\een}{\end{enumerate}}
\def\bes{\begin{equation*}}
\def\ees{\end{equation*}}
\def\bead{\begin{aligned}}
\def\eead{\end{aligned}}

\renewcommand{\(}{\left(}
\renewcommand{\)}{\right)}
\def\bmat{\left(\begin{matrix}}
\def\emat{\end{matrix}\right)}

\def\cL{{\cal L}}
\def\cO{{\cal O}}
\def\p{{\mathbf p}}

\usepackage{mfirstuc} % to uppercase the name
\newcommand{\addReviewer}[2]{
  \expandafter\newcommand\csname #1\endcsname[1]{{\textbf{ \color{#2} \capitalisewords{#1}:\,##1}}}
  \expandafter\newcommand\csname #1cor\endcsname[2]{{\color{#2} \capitalisewords{#1}:\,\st{##1}{\textbf{##2}}}}
  \expandafter\newcommand\csname #1color\endcsname{#2}
  \expandafter\newcommand\csname #1todo\endcsname[1]{{\todo[inline,color=white!70!#2, caption={}]{\textbf{\capitalisewords{#1}}: ##1}}}
}

\usepackage{soul,color}
\definecolor{chromeyellow}{rgb}{1.0, 0.65, 0.0}

\begin{document}

\title{Dark Matter in A Mirror Solution to the Strong CP Problem}

\author{Quentin Bonnefoy}
\email{q.bonnefoy@berkeley.edu}
\affiliation{Berkeley Center for Theoretical Physics, Department of Physics, University of California, Berkeley, CA 94720, USA}
\affiliation{Theoretical Physics Group, Lawrence Berkeley National Laboratory, Berkeley, CA 94720, USA}

\author{Lawrence Hall}
\email{ljh@berkeley.edu}
\affiliation{Berkeley Center for Theoretical Physics, Department of Physics, University of California, Berkeley, CA 94720, USA}
\affiliation{Theoretical Physics Group, Lawrence Berkeley National Laboratory, Berkeley, CA 94720, USA}

\author{Claudio Andrea Manzari}
\email{camanzari@lbl.gov}
\affiliation{Berkeley Center for Theoretical Physics, Department of Physics, University of California, Berkeley, CA 94720, USA}
\affiliation{Theoretical Physics Group, Lawrence Berkeley National Laboratory, Berkeley, CA 94720, USA}

\author{Amara McCune}
\email{amara@physics.ucsb.edu}
\affiliation{Berkeley Center for Theoretical Physics, Department of Physics, University of California, Berkeley, CA 94720, USA}
\affiliation{Theoretical Physics Group, Lawrence Berkeley National Laboratory, Berkeley, CA 94720, USA}
\affiliation{Department of Physics, University of California, Santa Barbara, CA 93106, USA}

\author{Christiane Scherb}
\email{cscherb@lbl.gov}
\affiliation{Berkeley Center for Theoretical Physics, Department of Physics, University of California, Berkeley, CA 94720, USA}
\affiliation{Theoretical Physics Group, Lawrence Berkeley National Laboratory, Berkeley, CA 94720, USA}

\begin{abstract}
We study thermal production of dark matter (DM) in a realization of the minimal models of Ref.~\cite{Bonnefoy:2023afx}, where parity is used to solve the strong CP problem by transforming the entire Standard Model (SM) into a mirror copy. Although the mirror electron $e^{\prime}$ is a good DM candidate, its viability is mired by the presence of the mirror up-quark $u^{\prime}$, whose abundance is intimately related to the $e^{\prime}$ abundance and must be suppressed. This can be achieved through a sequential freeze-in mechanism, where mirror photons are first produced from SM gluons, and then the mirror photons produce $e'$. After computing the details of this double freeze-in, we discuss the allowed parameter space of the model, which lies at the threshold of experimental observations. We find that this origin of $e'$ DM requires a low reheating temperature after inflation and is consistent with the baryon asymmetry arising from leptogenesis, providing mirror neutrinos have a significant degeneracy. Finally, we show that this $e'$ DM is not compatible with Higgs Parity, the simplest scheme with exact parity, unless SM parameters deviate significantly from their central values or the minimal model is extended.     
\end{abstract}

\maketitle

%%%%%%%%%%%%%%%%%%%%%%%%%%%%%%%%%%%%%%%
\section{Introduction}

The nature of DM and the strong CP problem are two of the most compelling puzzles of particle physics. In the last decades, cosmological observations of the cosmic microwave background (CMB), distant supernovae, large samples of galaxy clusters, and baryon acoustic oscillations have firmly established a standard cosmological model in which DM accounts for about 85\% of the matter content of the Universe, and about 27\% of the global energy budget~\cite{Planck:2015fie}. Another puzzle stems from the observational absence of the neutron electric dipole moment, $d_n < 10^{-26}\; \rm{e\cdot cm}$~\cite{Pendlebury:2015lrz}, which constrains the amount of CP violation in the strong interactions. (C stands for charge conjugation, P for spacetime parity and CP for their combination.) In the QCD Lagrangian, there is only one CP violating term~\cite{tHooft:1976rip},
\begin{equation}
    \mathcal{L} \supset \bar\theta_{\rm QCD}\frac{g_s^2}{32\pi^2} G^a_{\mu\nu}\tilde{G}^{a,\mu\nu}\,,
\end{equation}
where $G_a^{\mu\nu}$ is the gluon field strength tensor, $g_s$ the strong coupling constant, $\tilde{G}_{a,\mu\nu} \equiv \frac{1}{2}\epsilon_{\mu\nu\alpha\beta}G_a^{\alpha\beta}$ and $\bar\theta_\text{QCD}$ is an angle $\in[0,2\pi]$. In the SM, $\bar\theta_{\rm QCD}$ combines the bare $\theta$-angle $\theta_{\rm QCD}$ with an anomalous contribution from the quark mass matrix $M$ into a flavor-invariant quantity, $\bar\theta_{\rm QCD} \equiv \theta_{\rm QCD} + \arg\det(M)$. The constraint above results in the upper limit $\bar\theta_{\rm QCD} \lesssim 10^{-10}$ \cite{Baluni:1978rf,Crewther:1979pi,Pospelov:1999mv}, so that $\bar\theta_{\rm QCD}$ is by far, and inexplicably, the smallest dimensionless parameter of the SM. This strong CP problem is made even more surprising by the fact that weak interactions violate CP with a phase of order unity.

Three approaches to this problem have received considerable attention in the literature: a massless colored fermion~\cite{Kaplan:1986ru,Banks:1994yg,PhysRevLett.114.141801,Agrawal:2017evu} which makes $\bar\theta_{\rm QCD}$ unphysical (and whose minimal incarnation is now excluded by lattice data~\cite{FLAG:2021npn}), spontaneously broken P or CP symmetries~\cite{Nelson:1983zb,Barr:1984qx,Babu:1988mw,Babu:1989rb} that fix $\bar\theta_{\rm QCD}=0$ in the UV and rely on its extremely suppressed renormalization \cite{Ellis:1978hq,Valenti:2021rdu,deVries:2021pzl,Valenti:2022uii,Hisano:2023izx}, and a spontaneously broken global anomalous symmetry à la Peccei-Quinn \cite{Peccei:1977hh,Peccei:1977ur} which relaxes $\bar\theta_{\rm QCD}$ to $0$ through the QCD-induced dynamics of the predicted QCD axion \cite{Weinberg:1977ma,Wilczek:1977pj}. The QCD axion additionally turns out to be a natural DM candidate \cite{Preskill:1982cy,Abbott:1982af,Dine:1982ah}. (See Ref.~\cite{DiLuzio:2020wdo} for a review on the QCD axion and a large set of references.)

It was recognized in the 1970s already that parity might solve the strong CP problem~\cite{Beg:1978mt, Mohapatra:1978fy}, and several models have been analyzed in the literature since then (see e.g.~\cite{Babu:1988mw,Babu:1989rb,Barr:1991qx,Lavoura:1997pq,Gu:2012in,DAgnolo:2015uqq,Kawamura:2018kut,Hall:2018let,Dunsky:2019api,Hall:2019qwx,Craig:2020bnv,Redi:2023kgu,Carrasco-Martinez:2023nit}). Those models require extending the gauge group and particle content of the SM. However, unlike the QCD axion, the presence of a good DM candidate is not guaranteed by a suitable choice of parameters and cosmological history, as P relates the additional couplings and masses to the SM ones so that there is little freedom to find an appropriate parameter space. Extra fields can be added to play the role of DM \cite{Gu:2012in,Kawamura:2018kut}, while  within minimal models the possibility that DM is made out of the mirror neutrinos or mirror electrons predicted by parity has been explored in Refs.~\cite{Dror:2020jzy, Dunsky:2020dhn} and \cite{Dunsky:2019api}, respectively. For mirror neutrinos, justifying the necessary small parameters requires extra ingredients, while only non-thermal production mechanisms were known to work for mirror electrons. In this work, we study whether it is possible to thermally produce DM within a minimal setup where, as with the axion, its mass and abundance can be computed in terms of parameters already required to solve the strong CP problem.

Concretely, we analyze the model proposed by some of us in Ref.~\cite{Bonnefoy:2023afx}, based on a mirror copy of the SM. In the process, we automatically explore other setups studied in the literature~\cite{Barr:1991qx,Dunsky:2019api}, to which our model reduces in the appropriate limit. Parity fixes almost all coefficients given the measured SM parameters, so that one can genuinely scan the full parameter space. Generically, the only DM candidate is the mirror electron $e'$, i.e. the new fermion paired with the SM electron by parity. We argue that production from a dark sector in thermal equilibrium with the SM is not allowed by experimental constraints. Moreover, the only viable thermal production mechanism from the SM bath, that can adequately suppress the dangerous relic abundance of mirror up quarks $u^{\prime}$, is (sequential) freeze-in\footnote{Freeze-in of mirror electron DM has been considered in Twin Higgs models~\cite{Koren:2019iuv}. There, neutral naturalness does not lead to dangerous exotic hadrons and freeze-in through kinetic mixing is allowed.}. Mirror up quarks are electrically neutral and colored and hence can bind into fractionally-charged exotic hadrons. We perform a thorough numerical analysis and identify the range of model parameters where one finds consistent $e'$ DM. This fixes all scales of the model within a few orders of magnitude, dramatically increasing the predictivity beyond the constraints from parity alone.

As we were completing this work, a different cosmological history for the same model was presented in Ref.~\cite{Redi:2023kgu}, where $e'$ DM is obtained through freeze-out and subsequent dilution, while fractionally-charged exotic hadrons are argued to be sufficiently suppressed. While the bounds on such hadrons \cite{Dunsky:2019api} are based on fluxes that have uncertainties \cite{Dunsky:2018mqs}, they appear to exclude this cosmological scenario by several orders of magnitude, calling into question its viability.
 
The paper is organized as follows. In Sec.~\ref{sec:TheModel}, we summarize the relevant features and mechanisms of the model proposed in Ref.~\cite{Bonnefoy:2023afx} to solve the strong CP problem. In Sec.~\ref{sec:Running}, we discuss the running of masses and gauge couplings in this model, while in Sec.~\ref{sec:KM} we study the induced kinetic mixing between the SM and mirror photon, which strongly impacts the direct detection (DD) prospects of the model. In Sec.~\ref{sec:DM} we explore whether this model provides a solution to the DM problem, and in Sec.~\ref{sec:Hparity} we comment on the compatibility between DM and the Higgs parity mechanism, which allows parity to be exact instead of softly broken in the UV without adding new degrees of freedom. Finally, we conclude in Sec.~\ref{sec:Conclusion}. In App.~\ref{app:MPhDistr}, we compute the freeze-in distribution of mirror photons, allowing us to compute the $e'$ DM abundance in Sec.~\ref{sec:DM}.

%%%%%%%%%%%%%%%%%%%%%%%%%%%%%%%%%%%%%%%
\section{Mirror Solutions to the Strong CP problem}\label{sec:TheModel}

Here, we summarize the model presented in Ref.~\cite{Bonnefoy:2023afx}.

The full SM gauge group is mirrored to $SU(3)\times SU(2)\times U(1)_Y \times SU(3)^{\prime}\times SU(2)^{\prime}\times U(1)^{\prime}$, and the matter content is doubled to include mirror copies of the fermion and Higgs fields. One set of particles has the usual SM quantum numbers under $SU(3)\times SU(2)\times U(1)_Y$ and is a singlet of $SU(3)^{\prime}\times SU(2)^{\prime}\times U(1)^{\prime}$, while the converse is true for the other set. Each of the two Higgs fields is responsible for the breaking of the electroweak sector of its own ``world". This setup can solve the strong CP problem when the two worlds are related via a $\mathbb{Z}_2$ symmetry composed with the usual action of P, so that the (C)P-odd $\theta$ angles of $SU(3)$ and $SU(3)^{\prime}$ satisfy the relation $\theta = - \theta^{\prime}$. In Ref.~\cite{Bonnefoy:2023afx}, it was shown that breaking the two $SU(3)$ gauge groups to their diagonal subgroup, which is identified with $SU(3)_\text{QCD}$, provides a solution to the strong CP problem, as one finds $\theta_\text{QCD} = \theta + \theta^{\prime}$. In the limit of exact or softly broken P in the UV, the two $\theta$ angles cancel completely. In the IR, despite P being broken, $\theta_\text{QCD}$ receives negligible corrections. 

 The mirror Higgs acquires a vacuum expectation value (VEV) $v^{\prime} \gg v$, where $v$ denotes the SM Higgs VEV, which breaks P spontaneously. A phenomenologically-viable vacuum is achieved either by a soft P breaking term in the tree-level scalar potential or by radiative corrections at the scale $v'$~\cite{Hall:2018let}. In the latter case, $v'$ cannot be chosen arbitrarily and depends on the other couplings of the model. We discuss it further in Sec.~\ref{sec:Hparity}. In the rest of this paper, we treat $v'$ as an independent parameter, which can for instance be achieved through the aforementioned soft breaking of parity. In either case, the theory does not address the hierarchy problem. Moreover, in this paper, we consider the simplest $SU(3)\times SU(3)^{\prime} \to SU(3)_{\rm QCD}$ breaking mechanism, provided by the VEV of a bifundamental scalar field $\Sigma$. Although the solution to the strong CP problem does not depend on the specific breaking mechanism (see Ref.~\cite{Bonnefoy:2023afx} for additional examples), the phenomenology of the model is sensitive to it. This will be commented on when necessary. 

The interest of mirror solutions to the strong CP problem stems from their simplicity and predictivity: since the gauge and Yukawa couplings are fixed by P, the only two free parameters with respect to the SM are $v'$ and $v_3$, the energy scale at which $SU(3)\times SU(3)^{\prime}$ is broken. The constraints on the parameter space of this setup from collider searches and from the requirement that it solves the strong CP problem are shown in Fig.~1 of Ref.~\cite{Bonnefoy:2023afx}. The allowed region roughly reads
\begin{align}
\label{params for CP}
\begin{split}
2\cdot 10^8\, {\rm GeV} \lesssim v^{\prime} \lesssim 10^{14}\, {\rm GeV}; \;\;\;\; v_3 \gtrsim 5 \, {\rm TeV}.
\end{split}
\end{align}

%%%%%%%%%%%%%%%%%%%%%%%%%%%%%%%%%%%%%%%
\section{Masses, couplings and their running}\label{sec:Running}

As alluded to above, parity forces the gauge and Yukawa couplings in the two worlds to be equal at scales above $v'$,
\beq
\label{parityRelations}
g_{G'}=g_G \ , \quad y_{f'}=y_f \ ,
\eeq
where $G=1,2,3$ denotes a simple factor of the SM gauge group of coupling $g_G$ and similarly for the mirror gauge couplings $g_{G'}$, while $y_f$ are the SM Yukawa couplings in the up, down and charged lepton sectors and similarly for the mirror Yukawa couplings $y_{f'}$. The breaking of parity will generate deviations from these relations below $v'$ in a calculable fashion. Knowing the precise values will be of extreme importance for the discussion of DM below, so we discuss here the renormalization group equations (RGEs) of our model.

For a choice of $v'$ and $v_3$, the values of the parameters at all scales and in both worlds can be found. Indeed, the solution to the RGEs is fully fixed by the following constraints: 
\begin{enumerate}
\item Matching to the measured SM parameters at low energies,
\item Continuity at $v_3$, except for 
\beq
\label{matchingRelationV3}
\frac{1}{\alpha_\text{QCD}}=\frac{1}{\alpha_s}+\frac{1}{\alpha_s'}
\eeq
(plus possible threshold corrections at loop level), where $\alpha^{(\prime)}_s=\frac{g_s^{(\prime)}{}^2}{4\pi}$ with $g_s^{(\prime)}$ the (mirror) color coupling constant above $v_3$,
\item Parity at $v'$, i.e. the relations of Eq.~\eqref{parityRelations}.
\end{enumerate}
For given $v'$ and $v_3$, these constraints constitute sufficient boundary data to fix all integration constants in the RGEs.
We focus on the one-loop RGEs \cite{Cheng:1973nv} and use the values in the modified minimal subtraction ($\rm \overline{MS}$) scheme given in Ref.~\cite{Huang:2020hdv}. We have checked that using the two-loop RGEs for the strong couplings affect the result at the percent level, yielding a smaller source of uncertainty than that on the SM up quark mass (see below). The same applies to the use of the pole mass instead of the $\rm \overline{MS}$ mass. In this paper, we use the latter everywhere. We also include the effect of the bifundamental scalar field $\Sigma$ on the RGEs. For simplicity, we assume that all components of that scalar field acquire masses of order $v_3$ (this can be achieved upon suitably choosing the parameters of its scalar potential).

The most straightforward case is when the lightest colored mirror fermion, namely the mirror up quark $u'$, is heavy. Specifically, when $m_{u'}\geq v_3$. In this case, the modification of the SM RGEs is minimal all the way to $v'$: One simply needs to replace $\alpha_\text{QCD}\to \alpha_s$ and add the contribution of three flavors of fundamental scalars to the $SU(3)$ running. One can then pick a random value of $\alpha_s(v_3)$ (which lies between\footnote{The lower bound directly follows from Eq.~\eqref{matchingRelationV3}, while the upper bound also takes into account the fact that the mirror gauge coupling runs faster at lower energies, since the mirror fermions are heavier. Therefore, since the two gauge couplings are equal at $v'$, $\alpha_s(v_3)<\alpha_s'(v_3)$.} $\alpha_\text{QCD}(v_3)$ and $2\alpha_\text{QCD}(v_3)$), run the SM world couplings to $v'$, fix the mirror world parameters at $v'$ and run them down to $v_3$ to check whether the relation of Eq.~\eqref{matchingRelationV3} holds. Spanning over all (or many) choices of $\alpha_s(v_3)$, one can identify the appropriate value.

\begin{figure*}[bt]
	\centering
	\includegraphics[width=0.5\textwidth]{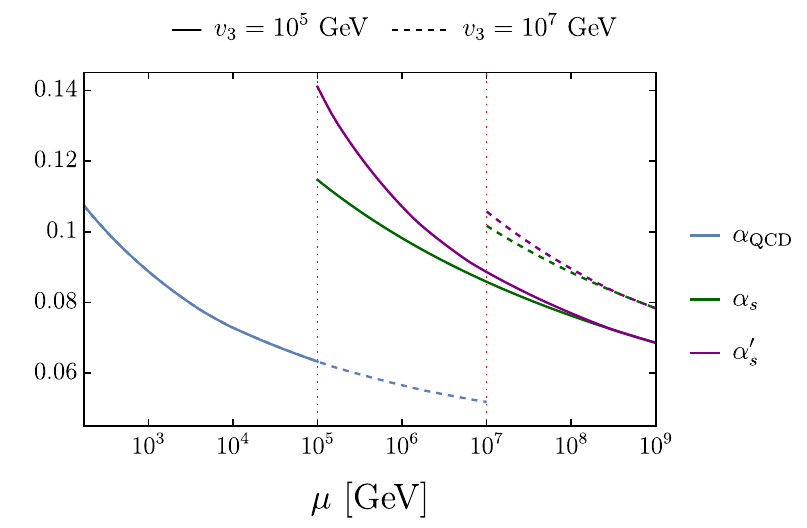}\qquad
 \includegraphics[width=0.44\textwidth]{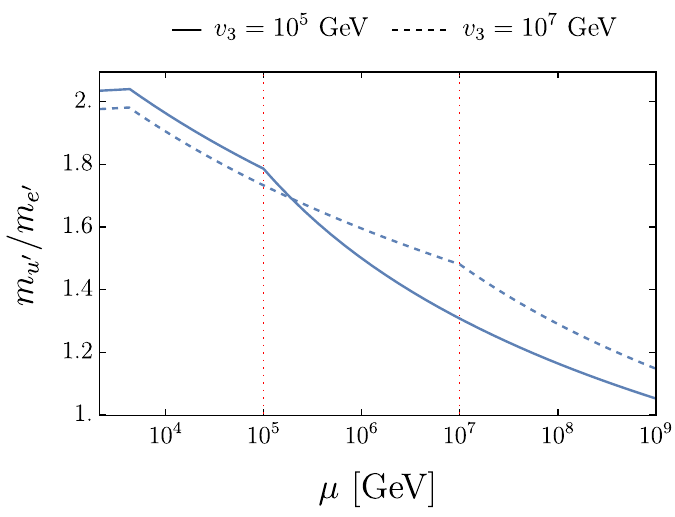}
	\caption{$\overline{\text{MS}}$ running parameters for $v'=10^9$ GeV and two choices of $v_3$, to which the red vertical lines correspond. Left panel: running gauge couplings of the color groups. Right panel: ratios of the $u'$ and $e'$ masses. The curves stop at $\mu=m_{e'}(\mu)$, and the running of $m_{u'}$ is frozen below $\mu=m_{u'}(\mu)$.}
	\label{fig:solsRGEs}
\end{figure*}

When $m_{u'}<v_3$, the situation is more intricate. The boundary values for the mirror quark masses depend on the SM quark masses, which depend on the running of the QCD coupling constant, which itself depends on the mirror quark masses and their contribution to the $\beta$ function. In this case, one needs to numerically solve the full system of RGEs of the two coupled mirror worlds. In practice, we implement this for all values of $m_{u'}/v_3$, and we cross-check the results with the previous method when applicable. We checked that our numerical precision is such that the constraints imposed by matching at $v_3$ and parity at $v'$ are satisfied at the sub-per-mille level. We illustrate some results of this procedure in Fig.~\ref{fig:solsRGEs}. In the right panel, we see that the loop correction to the ratio of $m_{u'}$ to $m_{e'}$, the mirror electron $e'$ mass, depends on $v_3$ and can be noticeable, which will prove crucial when we discuss DM later on.

The outputs of this procedure are the running gauge couplings and fermion masses. They are used as inputs in a second step in which we compute the running kinetic mixing and Higgs quartic, which we discuss below. The RGEs of the gauge couplings and of the Higgs potential have been studied using similar techniques in Ref.~\cite{Redi:2023kgu}.

A further comment on the input value for $m_u$ is in order. In the following, we find the DM production to be extremely sensitive to $m_{u'}/m_{e'}$, whose value is obtained from the low energy determination of the SM up-quark and electron masses, as described above. The electron mass is known with the astonishing precision of $\sim 0.1$ ppb, while the PDG reports $m_u(2\, \rm{GeV}) = 2.16^{+0.49}_{-0.26}\; \rm{MeV}$~\cite{ParticleDataGroup:2022pth}. This is the largest source of uncertainty in our result, as discussed in Sec. \ref{sec:DM}. The central value is used for the right panel of Fig.~\ref{fig:solsRGEs}. Note, that a recent combination of lattice determinations of $m_u$ reports a smaller uncertainty of $\sim4\%$~\cite{FLAG:2021npn}, but is not without controversy and remains under active investigation.

%%%%%%%%%%%%%%%%%%%%%%%%%%%%%%%%%%%%%%%
\section{Kinetic Mixing}\label{sec:KM}

Parity is compatible with a kinetic mixing between the gauge boson of $U(1)_Y$ and its mirror copy,
\begin{equation}
    \mathcal{L} \supset \epsilon F^{\mu\nu}F^{\prime}_{\mu\nu}\,.
\label{eq:lagKM}
\end{equation}
This term was not relevant for the solution to the strong CP problem in Ref.~\cite{Bonnefoy:2023afx}, but it constitutes an important source of cosmological constraints on the parameter space of the model. Indeed, once this term is introduced, all fermions charged under $U(1)^{\prime}$ become charged also under $U(1)_Y$. If this is the case for DM, very stringent constraints on its charge from direct detection experiments apply. Anticipating the discussion in Sec.~\ref{sec:DM}, the mirror electron is the only DM candidate of the model and kinetic mixing plays an important role in assessing its viability.

Removing the tree-level contribution to the kinetic mixing is a reasonable assumption. For instance, in UV completions where one $U(1)$ belongs to a larger gauge group, the interaction term in Eq.~\eqref{eq:lagKM} is not allowed by gauge invariance. Therefore, it vanishes above the scale at which the two $U(1)$ gauge theories emerge. However, it will be generated at loop-level below that scale. In the current setup, the first loop-level contribution to this term arises at 4-loop~\cite{Dunsky:2019api}, and could be further suppressed by an appropriate arrangement of the energy scales of the model, which we will discuss next. Nevertheless, due to the tight constraints, it is worth studying in detail the 4-loop contribution to $\epsilon$, as it depends only on the two free parameters of the model, $v^{\prime}$ and $v_3$.

\begin{figure}[h!]
	\centering
	\includegraphics[width=0.4\textwidth]{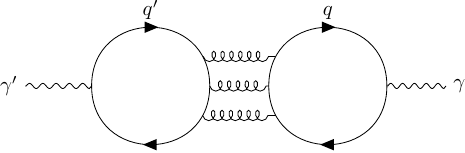}
	\caption{4-loop diagram contributing to kinetic mixing.}
	\label{fig:4loopKM}
\end{figure}

The leading diagram is shown in Fig.~\ref{fig:4loopKM}. As noted in Ref.~\cite{Dunsky:2019api}, the renormalization group equation of the kinetic mixing parameter can be read off from the four-loop beta function of QCD~\cite{vanRitbergen:1997va}, yielding
\begin{align}
\frac{{\rm d\epsilon}}{{\rm dln}\mu} = \frac{e^2g_s^6}{(4\pi)^8} \left(  - \frac{1760}{27} + \frac{1280}{9}\zeta(3) \right) \sum_{i j}q_i q'_j \ .
\label{eq:4loop}
\end{align}
It is clear that this contribution is only present below $v_3$, as above that scale the internal gluons do not couple simultaneously to SM and mirror quarks. Above $v_3$, we need higher-loop diagrams to connect $\gamma$ and $\gamma^{\prime}$, which are further suppressed.
\begin{figure}[h!]
	\centering
	\includegraphics[width=0.4\textwidth]{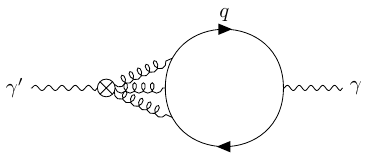}
	\caption{Diagram contributing to kinetic mixing below the lightest mirror quark mass.}
	\label{fig:epsIR}
\end{figure}
Moreover, below their masses, the mirror quarks are integrated out at one-loop and the three-loop diagram of Fig.~\ref{fig:epsIR} arises, where the effective vertex on the left connects one mirror photon to three gluons\footnote{There is no effective operator connecting one mirror photon to any combination of two gauge fields, due to symmetry (see the discussion on the ``$X^3$" class of operators in section 5 of \cite{Grzadkowski:2010es}).} and is therefore suppressed by $1/m_{u'}^4$ (note that, due to parity, $m_{u'}$ is the smallest mass scale among the UV fields). Dimensional analysis dictates that such contributions must come with an extra factor $\sim (\frac{p}{m_{u'}})^4$, where $p$ is the largest momentum flowing in the quark loops, while one retains the $e^2 g_S^6$ and four-loop suppression (one for the one-loop matching at $\mu\sim m_{u'}$ and three for the diagram in the IR theory). Therefore, one can safely neglect the running below $m_{u^{\prime}}$. In conclusion, 4-loop contributions to the kinetic mixing are relevant below the scale $v_3$ and when there is at least one mirror quark below that scale. 
%

%%%%%%%%%%%%%%%%%%%%%%%%%%%%%%%%%%%%%%%
\section{Dark Matter}\label{sec:DM}

Given the unbroken mirror QED symmetry of the model of Ref.~\cite{Bonnefoy:2023afx}, it is natural to expect the presence of a good DM candidate. However, it turns out that this happens only in a specific region of parameter space and through a non-standard production mechanism. This is the topic of this section.

Although little is known about the particle nature of DM, a good DM candidate must:
\begin{enumerate}
    \item Have a relic abundance today that matches the observed value $\Omega_{DM} = \rho_{DM} / \rho_{\rm crit} = 0.265(7)$, where $\rho_{\rm crit} = 8.5(1)\times10^{-30}\, {\rm g\,cm^{-3}} $ is the critical density of the Universe~\cite{ParticleDataGroup:2022pth},
    \item Be cosmologically stable so as to agree with current experimental observations~\cite{Audren:2014bca,Baring:2015sza,Mambrini:2015sia,Slatyer:2016qyl},
    \item Become non-relativistic well before matter-radiation equality, as required by the formation of large scale structures in the Universe~\cite{Benson_2012,Lovell:2013ola,Kennedy:2013uta},
    \item Have zero, or very small, electromagnetic charge, as required by searches for stable charged particles~\cite{McDermott:2010pa,Sanchez-Salcedo:2010gfa},
    \item Have limited self-interactions, constrained by the observed DM halo profiles, cluster collisions and the CMB spectrum~\cite{Agrawal:2016quu}.
\end{enumerate}
In the present model, all mirror particles are electrically neutral, up to kinetic mixing. Through mirror Yukawa and electroweak couplings fixed by parity, the mirror particles have decay channels similar to their SM counterparts, hence the only stable massive particles are the mirror electron $e'$ and up quark $u'$. They cannot decay to the SM because of the unbroken mirror electromagnetic charges that they carry. The mirror photon remains massless and contributes to dark radiation (by an amount far below current limits in the scheme to be described), whereas mirror neutrinos are heavy but quickly decay to the SM unless further model building is invoked, as we discuss below. Finally, the physical components of the bifundamental scalar field $\Sigma$ which breaks the color groups are generically of mass $\sim v_3$ and quickly decay to quarks and gluons \cite{Bai:2018jsr}.

Being the lightest stable mirror fermion, and charged only under mirror gauge groups, the mirror electron is the best DM candidate in this class of models. However, in Sec.~\ref{sec:Running} we have seen that the $u'$ has a mass close to the $e^{\prime}$, and so its abundance has to also be carefully considered. This will turn out to be of utmost importance for the viability of a DM production mechanism, given the bounds on the $u^{\prime}$ relic density. Another important constraint on the cosmology stems from the fact that the potential of $\Sigma$ displays an accidental $\mathbb{Z}_3$ symmetry, leading to domain walls (DWs) if the universe is reheated above the phase transition temperature $\sim v_3$. (This symmetry appears to be accidental to all mass dimensions, preventing one from introducing a bias to collapse the domain walls.) However, over most of the parameter space, this requirement is  weaker than the one associated with the relic density of $u'$.

\subsection{Bounds on mirror quarks}\label{sec:u'Bounds}

An extensive discussion of the cosmological history of the mirror quarks in these models can be found in Ref.~\cite{Dunsky:2019api}. The heavier mirror quarks decay into the lighter ones, while the latter hadronize after the QCD phase transition, forming bound states by combining with other colored particles. Rearrangements mediated by scattering processes then lead to the presence of two kinds of exotic bound states:
\begin{enumerate}
\item Hadrons made of $u^{\prime}$ and SM quarks: $ u^{\prime}qq$, $u^{\prime}u^{\prime}q$, $u^{\prime}\bar{q}$. We denote these by $h^{\prime}$. 
\item Hadrons made of three $u^{\prime}$.
\end{enumerate}
The abundance of these states, relative to the abundance of $u^{\prime}$, has been estimated in Ref.~\cite{Dunsky:2019api},
\begin{equation}
\begin{split}
Y_{u'u'u'} &\simeq Y_{u'} \times
\begin{cases}
Y_{u'} / Y_{\rm crit} & Y_{u'} < Y_{\rm crit}\\
1 & Y_{u'} >Y_{\rm crit}
\end{cases}, \\
Y_{h'} &\simeq Y_{u'} \times
\begin{cases}
1 & Y_{u'} < Y_{\rm crit}\\
 Y_{\rm crit} / Y_{u'}& Y_{u'} >Y_{\rm crit}
\end{cases},  \\
Y_{\rm crit} &\simeq 2\times 10^{-13}\; \big(\frac{m_{u^{\prime}}}{\rm GeV}\big)^2 \; Y_{\rm DM} \ .
\label{eq:Ycrit}
\end{split}
\end{equation}
While bound states made of only $u^{\prime}$ can in principle be a component of DM, the abundance of $h^{\prime}$ is strongly constrained by searches for nuclear and electric recoil at deep underground detectors~\cite{CDMS-II:2009ktb, XENON:2018voc,Essig:2017kqs}, and by tracks of ionizing particles in bulk matter, on earth~\cite{Majorana:2018gib,MACRO:2004iiu,Kajino_1984,Alekseev:1983hpa} as well as in meteorites~\cite{Jones:1989cq}. Again we refer the reader to the discussion in Ref.~\cite{Dunsky:2019api}.

Collider bounds on fractionally charged heavy stable states give $m_{u^{\prime}} \gtrsim 1.5\, {\rm TeV}$~\cite{Bonnefoy:2023afx}, while bounds from higher dimensional operators that can spoil the solution to the strong CP problem imply $m_{u^{\prime}} \lesssim 10^6\; {\rm TeV}$. In this mass range, one obtains $Y_{h^{\prime}} \lesssim [10^{-14}-10^{-8}]\; Y_{\rm DM}$ from the bounds of the MACRO~\cite{MACRO:2004iiu}, ICRR~\cite{Kajino_1984} and Baksan experiments~\cite{Alekseev:1983hpa}, as well as $Y_{h^{\prime}} \lesssim m_{u^{\prime}}/ {\rm GeV}\, 10^{-15}\; Y_{\rm DM}$ from searches of ionizing tracks in the Hoba meteorite~\cite{Jones:1989cq}. The lower bound on $m_{u'}$ and the definition of $Y_{\rm crit}$ imply that $Y_{\rm crit} \gtrsim 4\times 10^{-7}\; Y_{\rm DM}$, hence $Y_{h'} < Y_{\rm crit}$ given the bounds above. This shows that the only viable scenario is 
\begin{equation}
    Y_{u^{\prime}} < Y_{\rm crit},\quad Y_{h^{\prime}}\simeq Y_{u^{\prime}}, \quad Y_{u^{\prime}u^{\prime}u^{\prime}} \simeq \frac{Y_{u^{\prime}}^2}{Y_{\rm crit}} \,.
\end{equation}
In the literature, it has been debated whether strongly interacting particles can reach terrestrial experiments such as MACRO, ICRR and Baksan, because of their interactions in the Earth's atmosphere and crust. However, even assuming a geometrical cross section for DM-nucleus scattering (which is the largest possible value for such a cross section\footnote{A larger cross section can be obtained through long range interaction or, in the case of composite DM, by the presence of resonances or threshold bound states~\cite{Digman:2019wdm}.}~\cite{Digman:2019wdm}), it has been shown that this is the case for neutral DM bound states~\cite{Goodman:1984dc} as well as charged ones~\cite{Dunsky:2019api}. The determination of the $h'$ flux in the galactic disk from its charged massive particle nature~\cite{Dunsky:2018mqs} is not straightforward, hence the systematic errors on the bounds for $Y_{h'}$ are difficult to estimate. Therefore, in the following, we show different exclusion scenarios. Nonetheless, the constraints are so stringent that, even taking a conservative stand, they play an important role in our results.

Concerning bound states made of only $u^{\prime}$, one finds from the relations above that $Y_{u'u'u'}\lesssim 2\times 10^{-10}\; Y_{\rm DM}$, and the DM fraction of fully mirror bound states is negligible.  

\subsection{Bounds on $e^{\prime}$ DM}

Since bound states of mirror quarks are constrained to provide only an extremely small contribution to DM, we are left with the possibility of mirror electron DM. We remind the reader that in this model the mirror electron is charged only under the mirror gauge groups $SU(2)^{\prime}\times U(1)^{\prime}$ and has a mass larger than $\sim 750\, {\rm GeV}$\footnote{This limit is obtained from the bound on the mirror quark masses which, due to parity, implies a bound on the mirror electron mass with a mild dependence on the parameters of the model.}. Before studying the production mechanism in the early universe, we discuss the bounds from DM direct and indirect searches.  

\begin{figure*}[tb]
	\centering
	\includegraphics[width=1.0\textwidth]{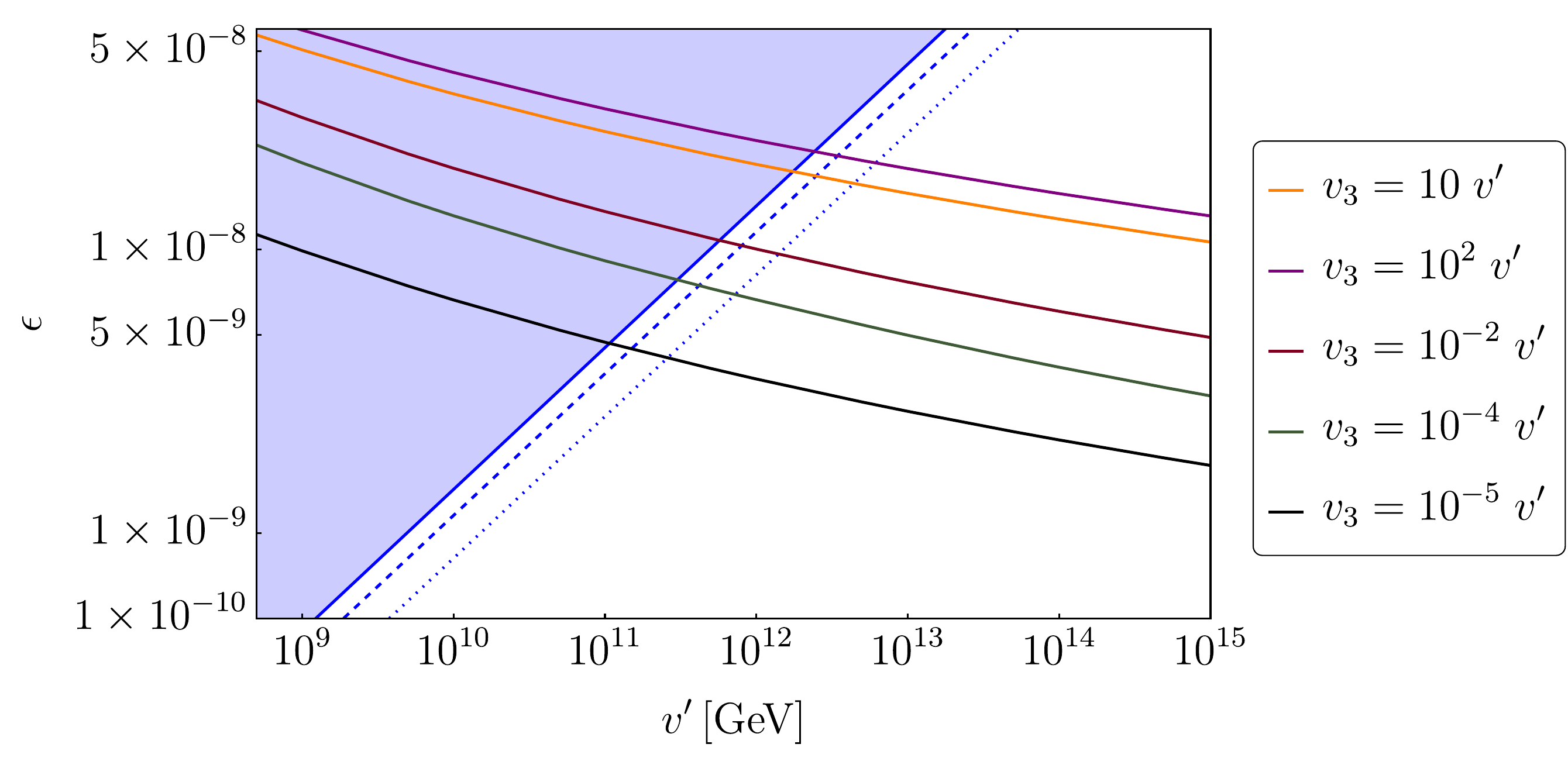}
	\caption{Limit on $\epsilon$ vs $v^{\prime}$ from DM direct searches (blue region) from XENON1T \cite{XENON:2018voc} (solid line), XENONnT \cite{XENON:2023cxc} (dashed line) and LUX-ZEPLIN \cite{LZ:2022lsv} (dotted line) together with predictions for different values of $v_3/v^{\prime}$.}
	\label{fig:epsvplimit}
\end{figure*}
Being coupled with the mirror photon, the $e^{\prime}$ experiences a long-range force. However, its mass is too heavy to appreciably self-scatter and disrupt the DM halo profile~\cite{Agrawal:2016quu} nor the spectrum of the CMB. On the other hand, the kinetic mixing discussed in Sec.~\ref{sec:KM} leads to an interaction with SM particles and can produce an observable signal. The strongest bound comes from searches for DM-nucleus scattering. The results of the experiment XENON1T~\cite{XENON:2018voc} can be recast~\cite{Dunsky:2019api} into the bound
\begin{equation}
    m_{e^{\prime}} > 10^6\, {\rm GeV} \bigg(\frac{\epsilon}{10^{-8}}\bigg)^2\,.
\end{equation}
Results from XENONnT \cite{XENON:2023cxc} and LUX-ZEPLIN (LZ) \cite{LZ:2022lsv} lead to similar bound, up to the respective replacement of $10^6$ by $1.5\times 10^6$ and $3\times 10^6$.
Since the mass of the mirror electron depends only on the VEV of the mirror Higgs with a very good precision, this limit translates into a bound on $\epsilon$ and $v^{\prime}$, as shown in Fig.~\ref{fig:epsvplimit}.
The same figure also shows the predictions of $\epsilon$ for different values of $v_3$, assuming no UV kinetic mixing as discussed in Sec.~\ref{sec:KM}. As anticipated there, the smaller $v_3$, the weaker the bound on $v^{\prime}$. For $v_3$ below the lightest mirror quark mass, the kinetic mixing is extremely suppressed and this bound disappears. For very low values of $v_3$ (above $m_{u^{\prime}}$), we find the lower limit $v^{\prime} \gtrsim 10^{11}\, {\rm GeV}$. This complements the other bounds on the parameters of the model. In fact, Ref.~\cite{Bonnefoy:2023afx} finds the bound $v^{\prime} \lesssim 10^{14}\, {\rm GeV}$ from higher-dimensional operators that can spoil the solution to the strong CP problem. Therefore, in the scenario $v_3 > m_{u^{\prime}}$, this leaves only 2-3 orders of magnitude for $v^{\prime}$.

\subsection{Freeze-out}\label{sec:FZO}

\begin{figure*}[bt]
	\centering
	\includegraphics[width=0.25\textwidth]{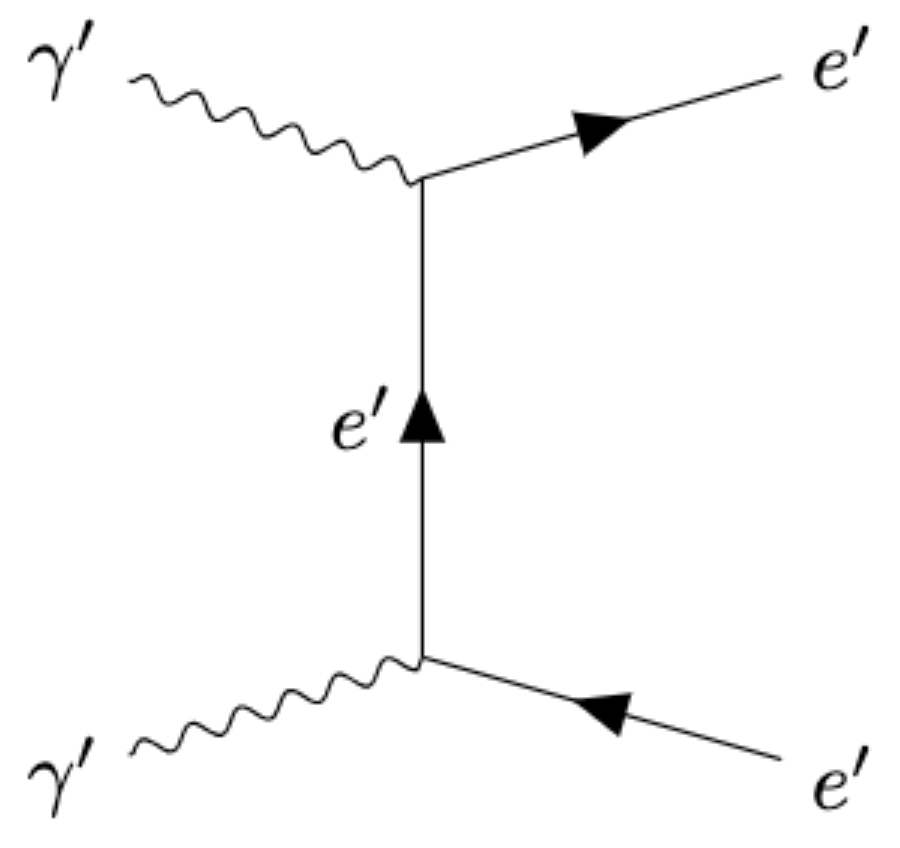}
    \includegraphics[width=0.25\textwidth]{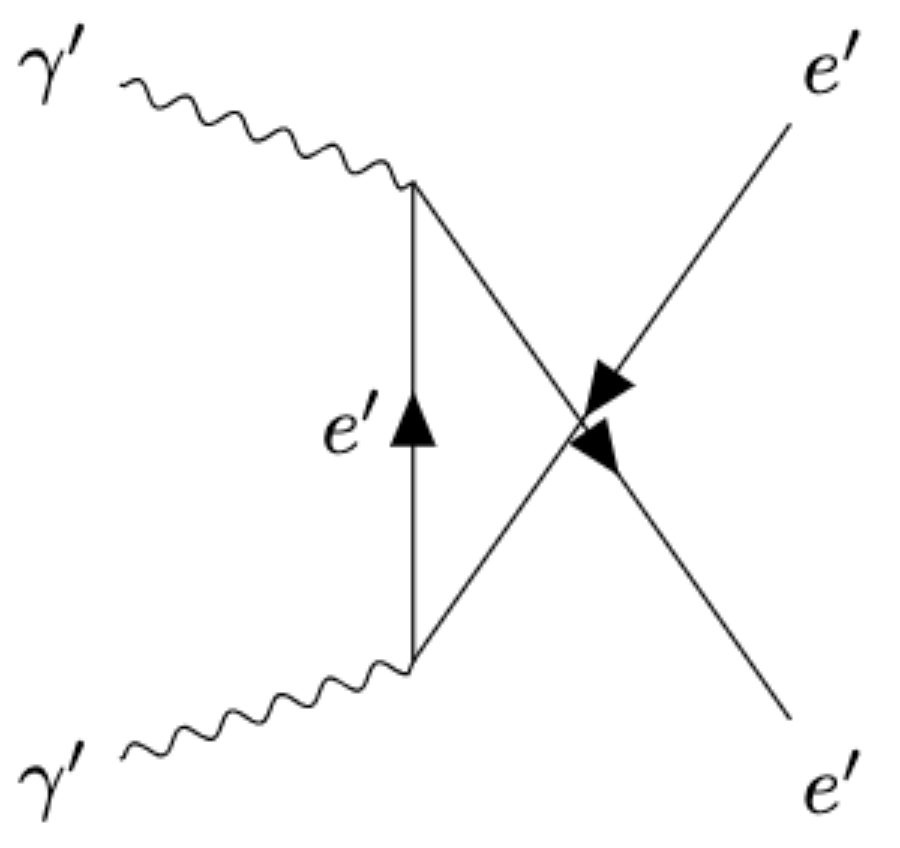}
    \includegraphics[width=0.36\textwidth]{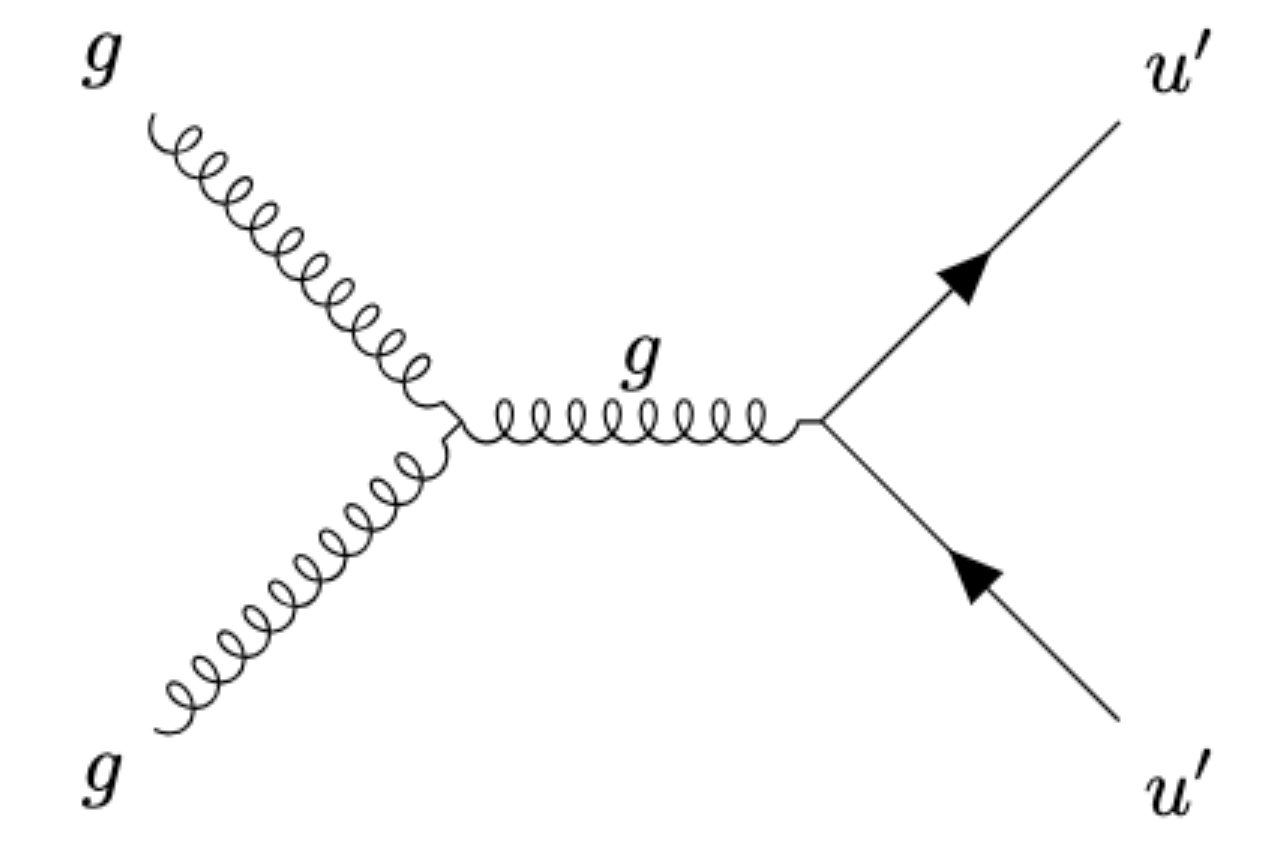}\\
    \includegraphics[width=0.25\textwidth]{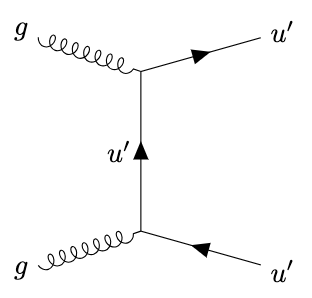}
    \includegraphics[width=0.25\textwidth]{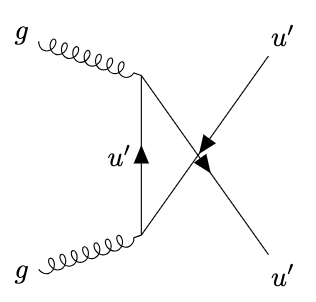}
    \includegraphics[width=0.34\textwidth]{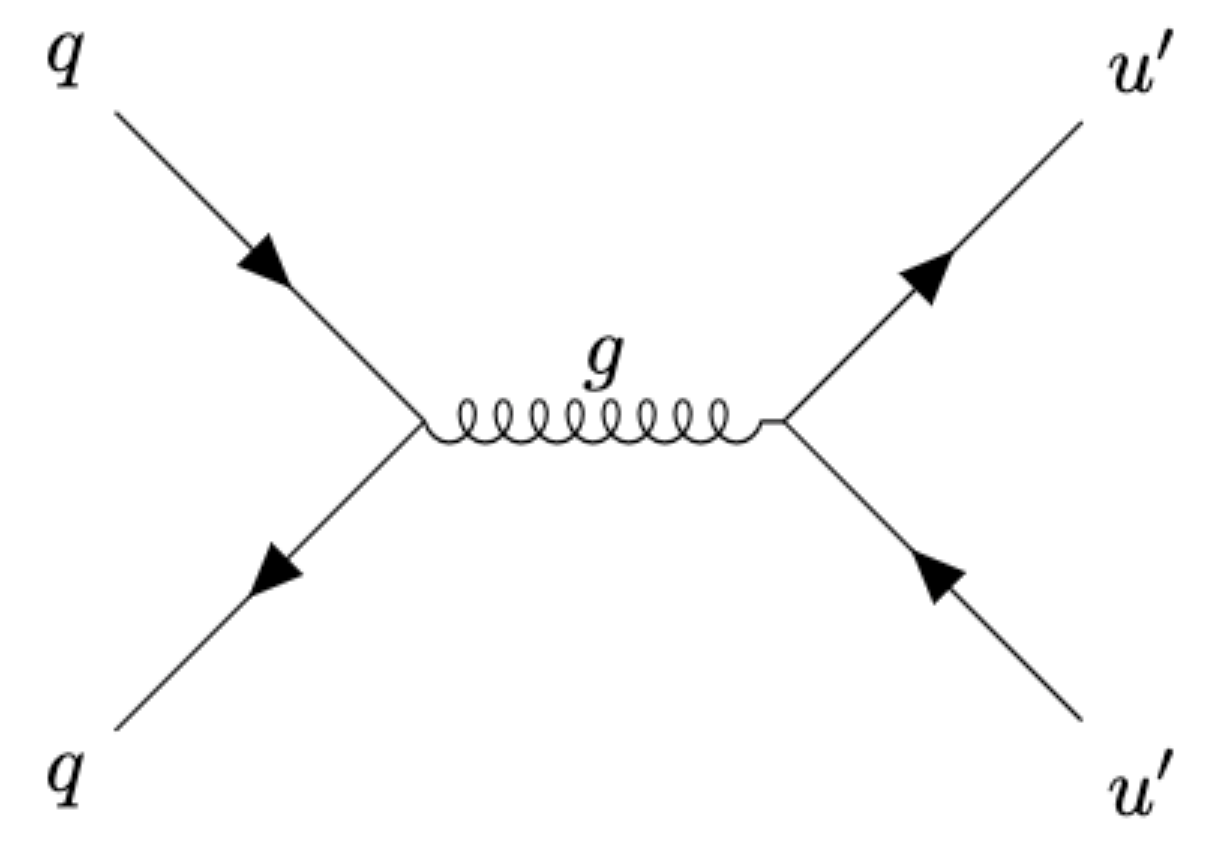}
	\caption{Most relevant tree-level diagrams for the production of $e^{\prime}$ and $u^{\prime}$ in the early universe when the temperature is smaller than $v_3$.}
	\label{fig:epupProduction}
\end{figure*}

The most relevant processes for the interaction of $e^{\prime}$ and $u^{\prime}$ with the thermal bath are shown in Fig~\ref{fig:epupProduction} when the temperature is smaller than $v_3$.
At temperatures higher than its mass, $e^{\prime}$ is efficiently produced by/annihilated into two mirror photons. In the freeze-out scenario, the resulting $e^{\prime}$ yield reads $Y_{e^{\prime}} \simeq Y_{\rm DM}\bigg(\frac{v^{\prime}}{10^8\; {\rm GeV}}\bigg)$ and the point in parameter space which reproduces the right DM abundance is already excluded by collider searches, as mentioned in Sec.~\ref{sec:TheModel}. Refs.~\cite{Dunsky:2019upk, Redi:2023kgu} studied the possibility of heavier $e'$, whose abundance is diluted trough the decay of the mirror neutrinos. 
However, we discussed the running of the mirror fermion masses in Sec.~\ref{sec:Running} and we saw that the ratio $m_{u^{\prime}}/m_{e^{\prime}}$ can be at most as large as $\sim 2$ for $m_{u^{\prime}}$ close to $v_3$. This implies that the abundance of $e^{\prime}$ and $u^{\prime}$ would be similar. Indeed, for $u^{\prime}$ the mirror photon channel is also active, but it is even dominated by the production/annihilation through gluons or SM quarks. For a temperature above $v_3$, the gluons become mirror gluons, and the bifundamental scalar $\Sigma$ is also in equilibrium and can annihilate into a $u'$ pair. It then appears to us that the bounds discussed in Sec.~\ref{sec:u'Bounds} are too strong and prevent the freeze-out scenario from being viable. Weakening the bounds on the $u'$ abundance by a few orders of magnitude to account for the large uncertainties is not sufficient to change the conclusion.

\subsection{Freeze-in}\label{sec:FZI}

We therefore need to assume that the mirror fermions are never in thermal equilibrium and are produced through a freeze-in mechanism. If $m_{u^{\prime}} > m_{e^{\prime}} \gg T$, a factor of a few in the mass leads to an exponential suppression in the abundances,  $Y_{u^{\prime}}/Y_{e^{\prime}}\sim \exp(\frac{m_{u^{\prime}}-m_{e^{\prime}}}{T})$, due to the Boltzmann suppression of the thermal bath particles which are energetic enough to produce $e^{\prime}$ and $u^{\prime}$.

In the following, we assume that the reheating temperature at the end of inflation, $T_R$, is the highest temperature $T_\text{max}$ reached in the cosmological history of the universe and is smaller than $m_{e^{\prime}}$. We are thus adding a new parameter to the model, whose cosmology is now determined by three quantities: $v^{\prime},\, v_3,\, T_R$. In addition, we need to assume that the inflaton does not directly decay to mirror fermions, or with a very reduced rate, and that it does not produce too many mirror photons either, as those would subsequently generate a population of mirror fermions\footnote{We note that an initial population of mirror photons is an interesting possibility, but we choose to focus here on the minimal, purely thermal DM production from the SM bath.}. None of this happens if we assume that the inflaton primarily decays to the SM sector, or that it mostly produces gluons and mirror gluons. The former is not in contradiction with exact parity due to the different electroweak VEVs in the two sectors, for instance if the inflaton couples to matter through a parity-symmetric Higgs portal \cite{Berezhiani:1995am}. For a $P$-odd inflaton, such a coupling and an inflaton VEV can generate a soft $P$ breaking leading to asymmetric electroweak scales in the two sectors, even though a severe tuning is needed. We elaborate on our assumptions regarding reheating in Sec.~\ref{sec:Inflation}.

\begin{figure}[h!]
	\centering
	\includegraphics[width=0.8\columnwidth]{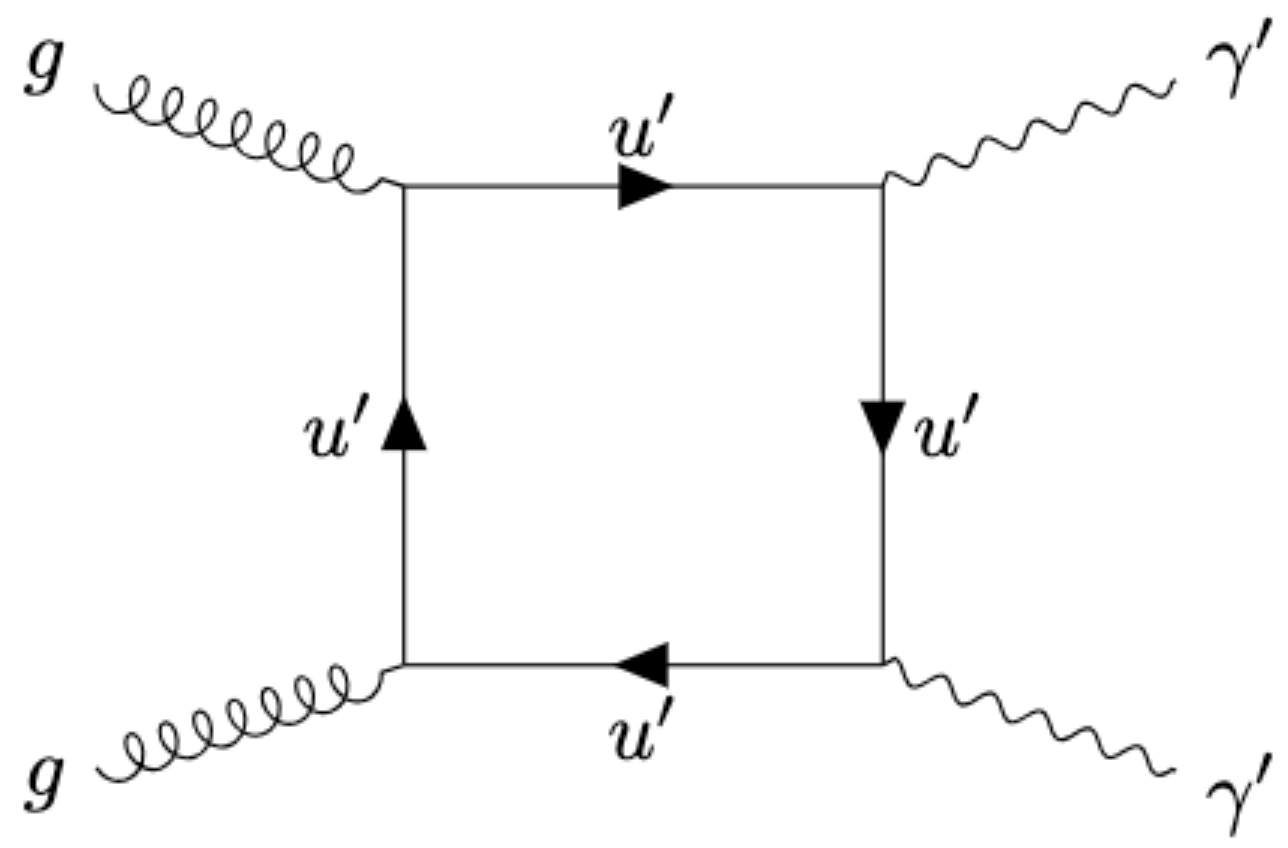}\\
    \includegraphics[width=0.8\columnwidth]{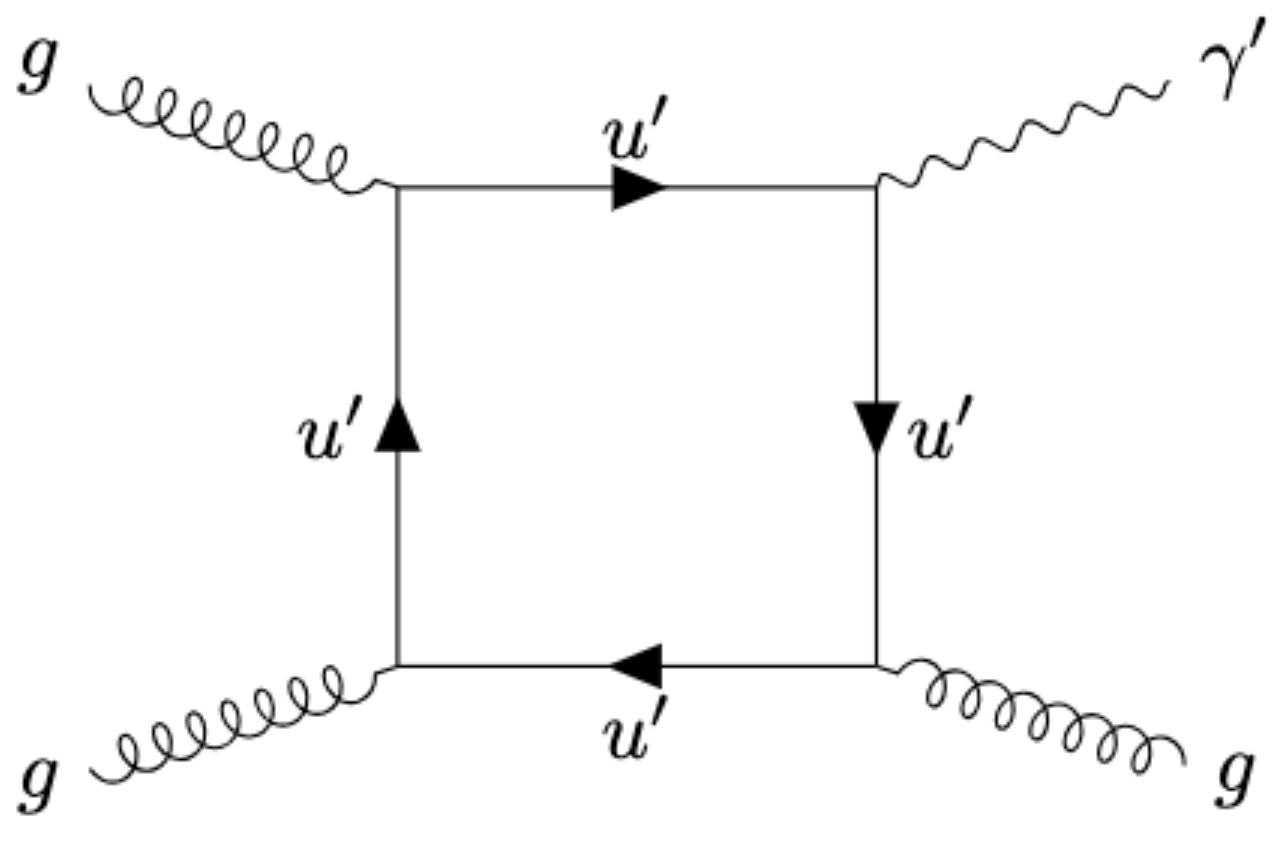}
	\caption{1-loop portal between SM gluons and  $\gamma^{\prime}$.}
	\label{fig:gggammagamma}
\end{figure}

In this setup, the only mirror species that can ever be in thermal equilibrium with the SM is the mirror photon, which, below $v_3$, interacts with the SM gluons through the 1-loop diagrams in Fig.~\ref{fig:gggammagamma}. Again, the relevant production processes for $e^{\prime}$ and $u^{\prime}$ below $v_3$ are shown in Fig.~\ref{fig:epupProduction}. Above $v_3$, $\Sigma$ thermalizes gluons and mirror gluons, so that the relevant diagrams become those of Fig.~\ref{fig:gggammagamma} and Fig.~\ref{fig:epupProduction} where gluons are replaced by mirror ones. 

\subsubsection{$\gamma^{\prime}$ in thermal equilibrium}

The cross sections for the processes in Fig.~\ref{fig:gggammagamma} read
\begin{equation}
    \sigma_{gg\to \gamma^{\prime}\gamma^{\prime}} \simeq 8\times 10^{-8}\,\frac{g_s^4 e^{\prime\, 4}s^3}{m_{u^{\prime}}^8} + O\bigg(\frac{s^4}{m_{u^{\prime}}^{10}}\bigg)\,,
    \label{higherDimOpCrossSection:ggg'g'}
\end{equation}
\begin{equation}
    \sigma_{gg\to g\gamma^{\prime}} \simeq 7.5\times 10^{-8}\,\frac{g_s^6 e^{\prime\, 2}s^3}{m_{u^{\prime}}^8} + O\bigg(\frac{s^4}{m_{u^{\prime}}^{10}}\bigg)\,,
    \label{higherDimOpCrossSection:gggg'}
\end{equation}
where $s = (p_1+p_2)^2$ and $p_{1,2}$ are the four-momenta of the external gluons. In the center of mass frame $s = 4E_1E_2$.

Following Ref.~\cite{GONDOLO1991145}, the rate for these processes is\footnote{\label{footnoteValidityXS}Note that we integrated over all energies in Eq.~\eqref{rateEquilibriumGammaPrime} although we used the simple formulae of Eqs.~\eqref{higherDimOpCrossSection:ggg'g'}-\eqref{higherDimOpCrossSection:gggg'} which are not valid at energies larger than $m_{u'}$. The exponential suppression above $E\sim T$ ensures that the result is not noticeably affected. For instance, accounting for resonant $u'\bar u'$ bound states formation~\cite{Pancheri:1992km,Barger:1987xg,Kahawala:2011pc,Kats:2012ym} yields an increase in the actual cross section that is comparable to that obtained from Eqs.~\eqref{higherDimOpCrossSection:ggg'g'}-\eqref{higherDimOpCrossSection:gggg'} when $s\sim m_{u'}^2$. We checked that this region and larger energies essentially do not contribute to the result.}
\begin{align}
\begin{split}
\Gamma_{\gamma^{\prime}} =& n_g \frac{\int \sigma v_{\rm Mol}\, e^{-E_1/T}e^{-E_2/T}d^3p_1d^3p_2}{\int e^{-E_1/T}e^{-E_2/T}d^3p_1d^3p_2}\\
&\simeq 1.5\times 10^{-3}\frac{(g_s^4 e^{\prime\, 4}+g_s^6 e^{\prime\,2})\,T^9}{m_{u^{\prime}}^8}\,,
\end{split}
\label{rateEquilibriumGammaPrime}
\end{align}
and the condition for having $\gamma^{\prime}$ in thermal equilibrium is
\begin{equation}
    \Gamma_{\gamma^{\prime}} \gtrsim H\, \simeq \frac{T^2}{M_P} \Rightarrow T_R\gtrsim  26\,{\rm GeV}\bigg(\frac{m_{u^{\prime}}}{{\rm TeV}}\bigg)^{8/7}\,. 
    \label{eq:Tequilibrium}
\end{equation}
The power of $m_{u'}$ indicates that $T_R$ grows faster than $m_{u'}$ (i.e., than $v'$) and tends to bring mirror fermions in thermal equilibrium, which is problematic as shown above. With this result, it is actually impossible to find a region of parameter space which gives the correct relic abundance for $e^{\prime}$. In fact, we find
\begin{equation}
    x_{e^{\prime}} \equiv  \frac{m_{e^{\prime}}}{T_R} \lesssim 38\, \frac{m_{e^{\prime}}}{m_{u^{\prime}}}\,\bigg(\frac{{\rm TeV}}{m_{u^{\prime}}}\bigg)^{1/7}\,.
\end{equation}
 As mentioned above, collider bounds require $m_{u^{\prime}}\gtrsim 1.5\, {\rm TeV}$, implying that $v'\geq 2\times 10^8$ GeV. For such values, we find numerically that the ratio $m_{e'}/m_{u'}$ is such that $x_e\lesssim 16$ all over the allowed parameter space, bringing the mirror electrons in thermal equilibrium with the mirror photons and overshooting the DM relic abundance by several orders of magnitude. (For instance, $v'=10^9$ GeV and $v_3=10^5$ GeV yield $x_e$ close to the maximal value, as can be seen in Fig.~\ref{fig:solsRGEs}.) We verified this estimate solving the Boltzmann equation numerically, and we indeed found that the mirror electrons reach thermal equilibrium. Importantly, the result holds even assuming a large uncertainty $\mathcal{O}(20\%)$ on $m_{u^{\prime}}$, in which case $x_{e^{\prime}}$ remains smaller than $\sim 22$.

\subsubsection{$\gamma^{\prime}$ out of equilibrium}

To reduce the abundance of $e^{\prime}$ and even further of $u^{\prime}$, $T_R$ needs to be lower than the limit in Eq.~\eqref{eq:Tequilibrium}. This implies that also $\gamma^{\prime}$ are out of equilibrium and DM is produced via a sequential freeze-in process~\cite{Hambye:2019dwd,Belanger:2020npe}. In App.~\ref{app:MPhDistr} we describe how to solve the Boltzmann equation for the momentum distribution of the mirror photons. We were able to solve it analytically in the limit where the number of degrees of freedom is constant and for high-energy mirror photons (i.e. $E_{\gamma'}\gg T$). These are consistent assumptions: the SM degrees of freedom are all in the bath until much lower temperatures than $T_R$ where most of the $e'$ are produced, and making $e'$ (such that $m_{e'} \gg T_R$ for freeze-in) requires highly energetic photons. Soft photons do not contribute to this process since they need to scatter with a very energetic one, whose number density is extremely suppressed. We checked the consistency of this argument, as detailed in App.~\ref{app:MPhDistr}.

\begin{figure}[h!]
	\centering
	\includegraphics[width=0.5\textwidth]{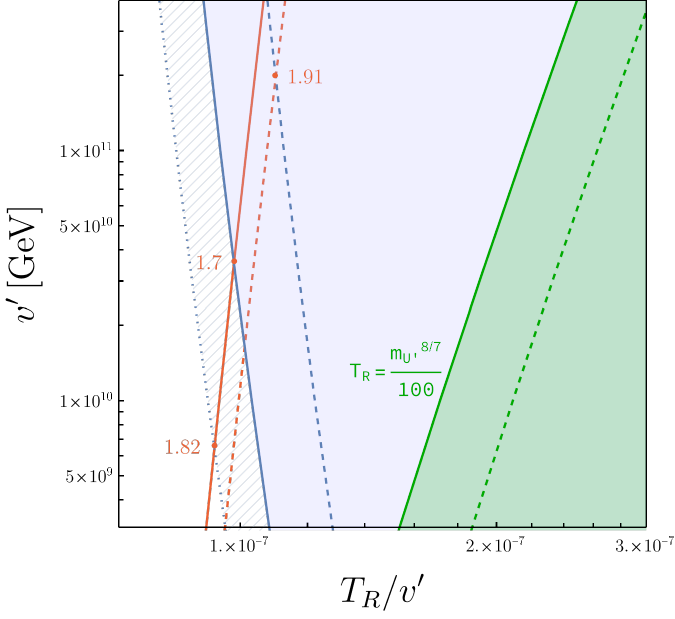}
	\caption{Values of $T_R$ and $v^{\prime}$ reproducing the correct DM relic abundance for $e^{\prime}$ (solid orange line) in the scenario $v_3 %= 10\, v^{\prime}
 >v'$. Requiring $Y_{h^{\prime}} < 10^{-8}\,Y_{\rm DM}$  ($Y_{h^{\prime}} < 10^{-12}\,Y_{\rm DM}$) rules out the blue region to the right of the solid blue (dotted blue) line. The region where the mirror photons distribution is thermal is shown in green. Dashed lines are analogous to solid ones, but using a 20\% higher value for $m_u$.}
	\label{fig:resultsDMlargev3}
\end{figure}

\begin{figure}[h!]
	\centering
	\includegraphics[width=0.5\textwidth]{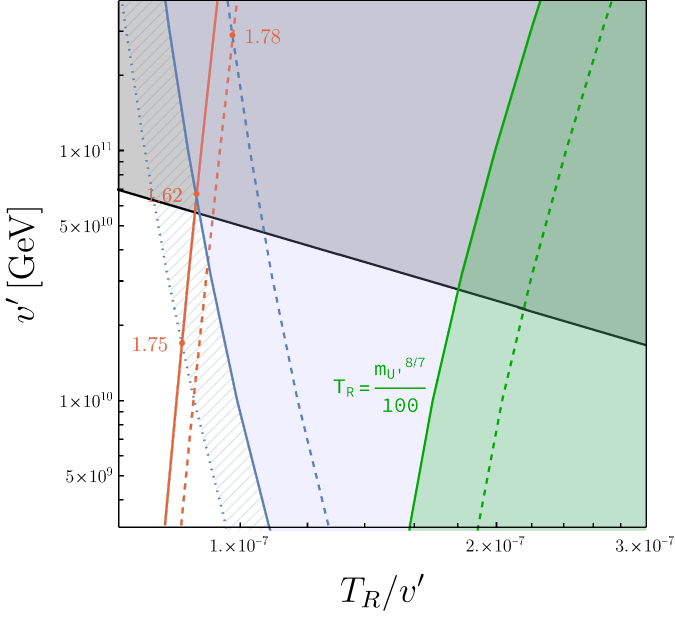}
	\caption{Values of $T_R$ and $v^{\prime}$ reproducing the correct DM relic abundance for $e^{\prime}$ (solid orange line) in the scenario $v_3 = 5\, {\rm TeV}$. Requiring $Y_{h^{\prime}} < 10^{-8}\,Y_{\rm DM}$  ($Y_{h^{\prime}} < 10^{-12}\,Y_{\rm DM}$) rules out the blue region to the right of the solid blue (dotted blue) line. The region where the mirror photons distribution is thermal is shown in green. Dashed lines are analogous to solid ones, but using a 20\% higher value for $m_u$. The gray region is such that $T_R>v_3$ and DWs are expected to form in the $\Sigma$ field.}
	\label{fig:resultsDMsmallv3}
\end{figure}

With the mirror photon distribution, we then numerically solve the Boltzmann equation for $e^{\prime}$ and $u^{\prime}$, making no further assumption and using the full non-thermal distributions of the mirror photons obtained in App.~\ref{app:MPhDistr}. Fig.~\ref{fig:resultsDMlargev3} and Fig.~\ref{fig:resultsDMsmallv3} show our results in the $v^{\prime} - T_R/v^{\prime}$ plane, for two benchmark cases:  when $v_3 > v^{\prime}$, and when $v_3$ has the smallest value allowed by collider searches ($v_3 = 5\, {\rm TeV}$), respectively. We discuss the extrapolation of intermediate cases later on, while for $v_3 > v^{\prime}$ the result is independent of $v_3$. (We fixed it to be $10 \, v'$ in our code to generate Fig.~\ref{fig:resultsDMlargev3}. The actual value of $v_3>v'$ only matters for kinetic mixing, whose RG running kicks in below $v_3$ as explained in Sec.~\ref{sec:KM}.) The points which provide the right yield of $e^{\prime}$ are shown with a solid orange line. The blue solid  and dotted lines show the experimental bounds on $u^{\prime}$, discussed in Sec.~\ref{sec:u'Bounds}. Given the uncertainty on these bounds, we show two benchmarks: $Y_{h^{\prime}} < 10^{-8}\,Y_{\rm DM}$ (solid blue)
and $Y_{h^{\prime}} < 10^{-12}\,Y_{\rm DM}$ (dotted blue). For comparison, the region where the mirror photons are in equilibrium is shown in green, confirming that it is incompatible with $e'$ DM. Moving to the right of the plot, the ratio $T_R/v^{\prime}$ increases as well as the abundances of both $e^{\prime}$ and $u^{\prime}$. A few values of $m_{u^{\prime}}/m_{e^{\prime}}$ are also shown along the DM line, from which one sees that the conservative bounds on the $u^{\prime}$ abundance give the constraint $m_{u^{\prime}}/m_{e^{\prime}} \gtrsim 1.6-1.7$ (with a small dependence on the value of $v_3$) to get a viable DM model. In the two figures, this translates to $v^{\prime}\lesssim 5\times 10^{10}\, {\rm GeV}$ and $T_R\lesssim 5\, {\rm TeV}$. Such a low reheating temperature raises the question of the baryogenesis mechanism at play in the early universe. We discuss this point in Sec.~\ref{sec:mirrorNus}. 

\begin{figure}[h!]
	\centering
	\includegraphics[width=0.5\textwidth]{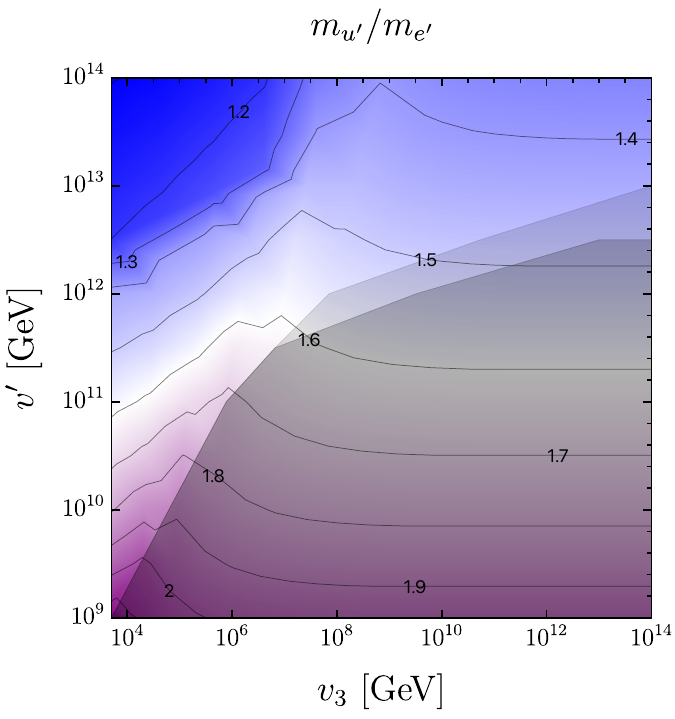}
	\caption{Ratio $m_{u'}/m_{e'}$ as a function of $v'$ and $v_3$. In the shaded region, $e'$ DM is excluded by the irreducible kinetic mixing and its impact on direct detection signals at Xenon1T (darker region) and LZ (lighter region).}
	\label{fig:muoverme}
\end{figure}

Anticipating the results of the numerical analysis for all values of $v'$ and $v_3$, the prediction for the kinetic mixing in Fig.~\ref{fig:epsvplimit} then suggests that models with $v_3 \gtrsim m_{u^{\prime}}$ do not contain a good DM candidate produced by freeze-in. This applies in particular to the models of Refs.~\cite{Barr:1991qx,Dunsky:2019api} (which can be obtained from ours in the $v_3\to \infty$ limit), and it can be seen in Fig.~\ref{fig:muoverme}, which shows the ratio $m_{u^{\prime}}/m_{e^{\prime}}$ as a function of $v_3$ and $v^{\prime}$. The observed trend is understood as follows: the ratio is larger for smaller $m_{u'}$, i.e. smaller $v^{\prime}$, as the $u^{\prime}$ feels the effect of a larger strong coupling constant which grows in the IR. Similarly, the smaller $v_3$, the larger its effect on the running of $m_{u^{\prime}}$ since strong couplings are typically larger above $v_3$, as illustrated in Fig.~\ref{fig:solsRGEs}. This effect competes with the fact that larger couplings make $m_u$ run faster, and that the RG running of the strong couplings above $v_3$ is steeper than the one below. Numerically, we find that the ratio $m_{u'}/m_{e'}$ is maximized for points where $m_{u^{\prime}} \sim v_3$, corresponding to the peaks in the contour plot of Fig.~\ref{fig:muoverme}. The bound from DM direct detection is also shown as a shaded region. It arises from the kinetic mixing between $\gamma$ and $\gamma^{\prime}$ as discussed in Sec.~\ref{sec:KM} and prevents the possibility of having DM for large values of $v_3$.
We can make the trend displayed in Fig.~\ref{fig:muoverme} sharper by running our RGEs and Boltzmann codes for all values of $v'$ and $v_3$. The result is shown in Fig.~\ref{fig:Hparity}, where we therefore identify a precise region of parameter space where the model explains the observed DM abundance while satisfying all other constraints:
\begin{equation}
\boxed{
\arraycolsep=1.4pt\def\arraystretch{1.5}
\begin{array}{c}
\text{\bf Parameters for Viable DM}\vspace{0.1cm}\\
\;\;\;10^9\, {\rm GeV} \lesssim v^{\prime}\lesssim 10^{11}\, {\rm GeV}\,,\;\;\;\vspace{0.1cm}\\
\;\;\;5\times 10^{3}\, {\rm GeV} \lesssim v_3 \lesssim 10^{-5}\; v'\, \,,\;\;\;\vspace{0.1cm}\\
 T_R \simeq  10^{-7} \, v' \, .
\end{array}
}
\label{eq:DMparam}
\end{equation}
In this parity solution to the strong CP problem, the mirror electron is a good DM candidate with a mass in the range $[5\, \rm{TeV} ,\, 250\, \rm{TeV}]$. We stress that the requirement of DM greatly reduces the large ranges for $v'$ and $v_3$ that solve the strong CP problem, shown in Eq.~\eqref{params for CP}.

We also investigated the impact of a large error, $\mathcal{O}(20\%)$, on the IR value of $m_u$. As commented in Sec.~\ref{sec:Running}, this is the major source of uncertainty in our result. The results of our numerical analysis are represented by the dashed blue lines in Figs.~\ref{fig:resultsDMlargev3} and~\ref{fig:resultsDMsmallv3}. These lines can be roughly reproduced upon rescaling the bounds (solid blue lines) through a $20\%$ shift in $T_R/v^{\prime}$. (This amounts to assuming a $\mathcal{O}(20\%)$ shift on $m_{u^{\prime}}$.) Note however that the DM line shifts to slightly higher $T_R$ (a $\sim 5\%$ effect) (dashed orange lines), due to the dependence of the mirror photon production cross section on $m_{u^{\prime}}$. Numerically, we find that increasing $m_{{u}^{\prime}}$ by $20\%$ extends the $e'$ DM region to $v^{\prime} \lesssim 2\times 10^{11}\, \rm{GeV}$ and $T_R \lesssim 25\, \rm{TeV}$. In addition, the lower bound on the ratio $m_{u^{\prime}}/m_{e^{\prime}}$ becomes larger as the bound on $Y_{u^{\prime}}$ becomes stronger. However, for low $v_3$ most of this region is excluded by the presence of $\Sigma$ domain walls.

On the other hand, lowering $m_{u}$ by $20\%$ gives stronger bounds, as expected, constraining $v^{\prime}$ between $10^9\, \rm{GeV}$ and $10^{10}\, \rm{GeV}$ and $T_R$ below $10^5\, \rm{GeV}$. Overall, this doesn't change the main picture of our result. The summary of results in the $(v_3,v')$ plane can be found in Fig.~\ref{fig:Hparity} below.

Finally, a non-thermal population of mirror photons remains until today, but it is far beyond observational prospects. Being more suppressed than thermal at $T_R$ and being subsequently diluted by the SM thresholds, it gives a very small contribution to dark radiation,  $\Delta N_\text{eff} \leq 7\times 10^{-6}$, which is saturated for the smallest reheating temperatures that allow $e'$ DM.

%%%%%%%%%%%%%%%%%%%%%%%%%%%%%%%%%%%%%%%

\subsection{Comments on Inflation and Reheating}\label{sec:Inflation}

Given the non trivial set of constraints which make DM viable in this model, we briefly comment on their impact on cosmological inflation and the subsequent reheating period. The following conditions are assumed in the predictive scenario for $e'$ production presented above.
\begin{itemize}
    \item At the end of reheating, when the universe enters a period of radiation domination, only the SM sector is populated and in thermal equilibrium at a temperature $T_R$ in the range [0.1 - 5] TeV. Higher $T_R$ would not generate an appropriate $e'$ relic density (see Figs.~\ref{fig:resultsDMlargev3} and \ref{fig:resultsDMsmallv3}).
    
    \item $T_R \leq \cO(0.1)\, m_{e'}$ and $m_{\phi} \lesssim m_{e^{\prime}}$, where $m_{\phi}$ is the mass of the inflaton. The former implies that our freeze-in calculation is applicable and that we do not freeze-out (hence overproduce) $e'$ and $u'$. The latter prevents the inflaton from directly decaying to $e'$ and $u'$, or its high-energy decay products from scattering and producing $e'$ or $u'$.

    \item During reheating, when the universe is dominated by the energy density of the inflaton field which is transferred to radiation, the temperature cannot be larger than $T_R$ to avoid populating the mirror sector\footnote{Calling $T_\text{max}$ the maximum temperature reached by the plasma during reheating, one may wonder if a qualitatively similar DM production could be performed if $m_{e'}>T_\text{max}>T_R$. In that case, $e'$ freeze-in happens during (inflaton) matter domination, and a large enough $e'$ relic density needs to be frozen-in in order to compensate the subsequent dilution. This implies that the Boltzmann factor $e^{-m_{e'}/T_\text{max}}$ should be increased with respect to the $e^{-m_{e'}/T_R}$ evaluated in Sec.~\ref{sec:FZI}. However, the correlated $e^{-m_{u'}/T_\text{max}}$ will also be increased, leading to $u'$ overproduction. In summary, $e'$ freeze-in at $T_\text{max}>T_R$ does not work in this model.}.
\end{itemize}
Furthermore, we require $T_R < v_3$ to avoid domain walls for $\Sigma$. It is clear that the possibilities for a viable inflationary model are strongly constrained.

A thorough analysis of inflation and reheating models is beyond the scope of this work; here we just note a few promising possibilities. The requirement that the maximal temperature reached during reheating is $T_R$ is fulfilled in the limit of instantaneous reheating, i.e. when the universe transitions from inflation to a phase of radiation domination in less than a Hubble time. For instance, this is achieved if  the inflaton potential makes an almost discontinuous change from being very flat to be very steep. Alternatively, one can deal with a smoother reheating if the temperature of the SM bath increases (or stays constant) throughout this phase, reaching a maximum at $T_R$. Scenarios of this kind, using dissipative processes other than decays, have been discussed in Ref.~\cite{Co:2020xaf}.

%%%%%%%%%%%%%%%%%%%%%%%%%%%%%%%%%%%%%%%
\section{Mirror neutrinos and leptogenesis} \label{sec:mirrorNus}

Finally, we discuss mirror neutrinos, which are electrically neutral.
In our minimal model, the neutrino masses must arise from Weinberg-type operators. Due to parity, one finds two independent coupling matrices $x_\nu,x'_\nu$, the former being symmetric and the latter hermitian \cite{Dror:2020jzy},
\beq
\bead
\cL_{\nu}&=\frac{x_{\nu,ij}}{2\Lambda}(HL_i)(HL_j)+\frac{x^*_{\nu,ij}}{2\Lambda}(H'L'_i)(H'L'_j)\\
&+\frac{x'_{\nu,ij}}{\Lambda}(HL_i)(H'L'_j) \ .
\eead
\label{eq:neutrinoMasses}
\eeq
Below $v'$, $H'$ is frozen to its vev and the mirror neutrinos acquire a mass matrix, $m_{\nu^{\prime}}$, and Yukawa coupling matrix to $LH$, $y_\nu$, of
\beq
m_{\nu^{\prime}} = x^*_{\nu} \frac{v^{\prime\, 2}}{\Lambda}, \hspace{0.5in} y_\nu = x'_\nu \frac{v^{\prime}}{\Lambda}\,.
\eeq
It is convenient to work in a basis where $m_{\nu^{\prime}}$ is real and diagonal. Below the electroweak scale a mass matrix arises also for the SM neutrinos: 
\beq
m_\nu = v^2\(\frac{m_{\nu'}}{v'{}^2}-y_\nu \, m_{\nu'}^{-1} \, y_\nu{}^T\)\,.
\eeq
The spontaneous breaking of parity at scale $v'$ leads to a ``direct" neutrino mass term proportional to $m_{\nu'}$ as well as a conventional ``seesaw" mass term proportional to $1/ m_{\nu'}$. If $x_{\nu,ij}$ and $x'_{\nu,ij}$ are comparable, as expected for example from approximate flavor symmetries, then the direct and seesaw contributions to the light neutrino masses will also be comparable, so that neutrino physics differs from that of just adding right-handed neutrinos to the SM. 

With $e'$ DM from sequential freeze-in, the cosmological effects of the mirror neutrinos are highly dependent on the coupling matrices $x, x'$ and the scale $\Lambda$.  For example, if the matrix elements of $x, x'$ are order unity, and $v'$ is of order $10^{10}$ GeV, the observed neutrino masses 
 require the scale $\Lambda$ to be of order $10^{15}$ GeV.  In this case the mirror neutrinos have masses of order $10^5$ GeV, well above the reheating temperature of $10^3$ GeV, so the mirror neutrinos are not made in the thermal bath and play no cosmological role. 

For other parameters the mirror neutrinos are light
enough to be produced at reheating, and they decay via the Yukawa coupling $y_\nu \, \nu' LH $ to $LH$ with decay rate
\beq
\Gamma_{\nu_i'\to \text{SM}}=\frac{1}{8\pi}\(y_\nu{}^\dagger y_\nu\)_{ii}m_{\nu_i'}\ .
\label{eq:nu'decay}
\eeq
Could such decays lead to the cosmological baryon asymmetry via thermal leptogenesis \cite{Fukugita:1986hr}? If there is no degeneracy among mirror neutrinos, the answer is no: for this case Ref.~\cite{Carrasco-Martinez:2023nit} finds that thermal leptogenesis requires $v^{\prime} > 10^{12}$ GeV to avoid fine tuning in the SM neutrino masses; Fig. \ref{fig:Hparity} shows this is inconsistent with $e'$ DM, even allowing for a large uncertainty in the up quark mass. Furthermore, thermal leptogenesis without $\nu'$ degeneracy requires the lightest $\nu'$ to be heavier than $10^9$ GeV, many orders of magnitude above $T_R$. 

However, it is well known that degeneracy among $\nu'$ produces the observed baryon asymmetry for much lower values of $m_{\nu'}$ \cite{Luty:1992un}. In this case, in theories with neutrino masses arising from (\ref{eq:neutrinoMasses}), leptogenesis can occur for lower values of $v'$: Fig. 5 of Ref.~\cite{Carrasco-Martinez:2023nit} shows that degeneracies in the range of $10^{-3}- 10^{-6}$, resulting from approximate flavor symmetries, gives successful leptogenesis throughout the allowed range of $v' \sim (10^9 - 10^{11})$ GeV required by $e'$ DM. 

An important feature of $e'$ DM from sequential freeze-in, relevant for leptogenesis, is that it requires a very low reheat temperature, $T_R \sim 10^{-7} v^{\prime} \sim (10^2 - 10^{4})$ GeV. Interestingly, this is above the electroweak scale, so any lepton asymmetry produced can be processed to a baryon asymmetry via electroweak sphalerons. A key question is the size of $m_{\nu_1'}$ relative to $T_R$. If $m_{\nu_1'}\gg T_R$, thermal production of $\nu'_1$, which occurs via the inverse of the decay process (\ref{eq:nu'decay}), will be highly Boltzmann suppressed, leading to a negligible lepton asymmetry. On the other hand, if $m_{\nu_1'}\lesssim T_R$ the produced lepton asymmetry will be strongly erased by rescatterings involving $\nu'_2$, which is degenerate with $\nu'_1$. Avoiding such strong washout requires reducing the Yukawa coupling coupling of $\nu'_2$ to the point that, at these low values of $m_{\nu_1'}$, the production of the asymmetry in $\nu'_1$ decays is too small, unless it is boosted by a degeneracy of at least $10^{-8}$. This exceeds the natural limit of degeneracy in this theory, $10^{-6}$, set by radiative corrections from the tau coupling \cite{Carrasco-Martinez:2023nit}. 

Thus the only possibility is that $\nu'_{1,2}$ have masses close to $T_R$, but sufficiently above that a near thermal abundance of $\nu'_1$ can be produced at reheating, while rescattering via virtual $\nu'_2$ leads to little washout of the asymmetry. We conclude that thermal leptogenesis may occur in the same minimal theory where $e'$ from sequential freeze in accounts for dark matter, but only if $\nu'_{1,2}$ are highly degenerate with mass several times larger than $T_R \sim  10^{-7} v^{\prime} \sim (10^2 - 10^{4})$ GeV.

Other possibilities for leptogenesis exist. The effective theory below $v'$ may contain the coupling $\phi \nu' \nu'$, allowing the inflaton $\phi$ to decay to mirror neutrinos as well as SM particles. Non-thermal leptogenesis then occurs in $\nu'$ decays before they thermalize. Even though strong washout may be avoided by having $m_{\nu'} \gg T_R$, some degeneracy among $\nu'$ is still required for a sufficient baryon asymmetry.  If the mirror neutrino is the lightest mirror fermion, inflaton decays to other mirror fermions may be kinematically forbidden, so that our previous calculations of the $e'$ and $u'$ abundance still applies. Another possibility is to augment the SM sector with gauge single fermions N and, by parity, SM$'$ is augmented by $N'$.  In this case thermal leptogenesis can result purely in the SM + $N$ sector via $N$ decay, as in conventional minimal leptogenesis. The low reheat temperature again requires significant $N$ degeneracy, but this is much less constrained by radiative corrections. The breaking of parity by $v'$ can lead to $N'$ coupling to $\nu'$ in pseudo-Dirac states much heavier than the reheat temperature. 

%%%%%%%%%%%%%%%%%%%%%%%%%%%%%%%%%%%%%%%
\section{Higgs Parity}\label{sec:Hparity}

\begin{figure*}[tb]
	\centering
	\includegraphics[width=1\textwidth]
 {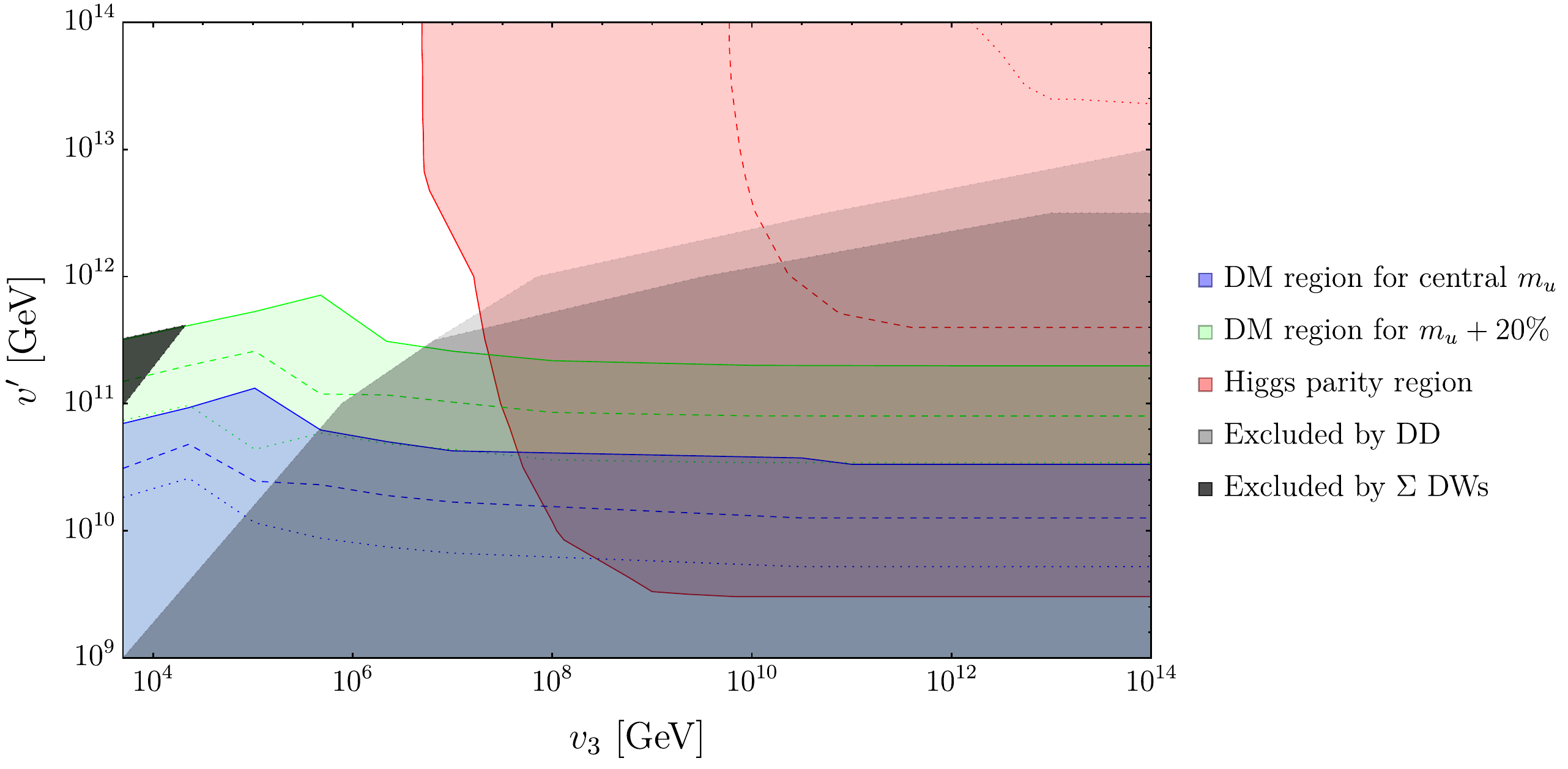}
	\caption{In the DM region, there exists a $T_R$ for which $e'$ constitutes all of DM while the $u'$ relics pass the experimental constraints. The solid, dashed and dotted curves assume that those constraints take the form $Y_{h'}< xY_\text{DM}$ with $x=10^{-8},10^{-10},10^{-12}$ respectively. The dark (light) gray-shaded region is excluded by direct detection searches for $e'$ DM by Xenon1T \cite{XENON:2018voc} (LZ \cite{LZ:2022lsv}), and the black-shaded one by the presence of $\Sigma$ domain walls. In the Higgs parity region, the Higgs quartic coupling vanishes at the scale $v'$ for some values of $\alpha_S$ and $m_t$ within their $2\sigma$ contours at $m_Z$. The dashed red line assumes central values for $\alpha_S$ and $m_t$, the solid line that they saturate their $2\sigma$ contours (in the direction of large $m_t$ and small $\alpha_S$), and the dotted one that $m_t$ takes its central value while $\alpha_S$ is increased to saturate its $2\sigma$ contour.}
	\label{fig:Hparity}
\end{figure*}

Models that implement a parity solution to the strong CP problem break parity either explicitly, via a soft breaking term in the potential, or spontaneously, with a vacuum having $v' \gg v$ stabilized by Coleman-Weinberg radiative corrections~\cite{Hall:2018let}. The latter mechanism, dubbed ``Higgs Parity", explains why the SM quartic coupling vanishes when evolved to a high energy scale, and implies that the new physics at this scale is that of parity restoration.

As described in Refs.~\cite{Dunsky:2019api,Dunsky:2019upk}, the tree-level parity symmetric potential for $H$ and $H'$ has an important feature: in the limit that $v \ll v'$ is imposed, an accidental $U(4)$ global symmetry emerges, with $\mathcal{H} =(H,H^{\prime})$ forming a fundamental representation so that an $SU(2)\times SU(2)^{\prime}$ subgroup is gauged.  When $\mathcal{H}$ gets a VEV, $\langle \mathcal{H} \rangle = (0,0,0,v^{\prime})$, the $U(4)$ is spontaneously broken to a $U(3)$, $H^{\prime}$ acquires a mass and at tree-level the SM Higgs arises as a massless Nambu-Goldstone boson. However, radiative corrections to the scalar potential (the leading contribution coming from the SM and mirror top quarks) break explicitly the $U(4)$ global symmetry, giving radiative contributions to the SM Higgs mass and quartic coupling. The large hierarchy $v'/v$ results mainly from fine-tuning and the SM Higgs quartic coupling $\lambda$ at the scale $v^{\prime}$ takes a small value. At lower energies, quantum corrections within the SM renormalize $\lambda$ so that it grows logarithmically. In this section, we discuss whether the condition $\lambda(v^{\prime}) \simeq -0.001$ \cite{Dunsky:2019api} is compatible with the parameter space giving rise to $e'$ DM.

The leading contributions to the RGE for $\lambda$ below the scale $v'$ are
\beq
\bead
\frac{d\lambda}{d\text{ln}\mu}=&\frac{24\lambda^2 + 
    12\lambda y_t^2 - 
    6 y_t^4}{16\pi^2}- 
 \frac{\lambda\(3 \alpha_1 + 
    9 \alpha_2\)}{4 \pi}\\ 
    &+ \frac{3}{8} \alpha_1^2 + 
   \frac{9}{8}\alpha_2^2 + 
   \frac{3}{4}\alpha_1\alpha_2\ \ ,
\eead
\eeq
where $y_t$ is the top Yukawa. The couplings $y_t, \alpha_1$ and $\alpha_2$ are computed at all scales, for any given $v_3$ and $v'$, as discussed in Sec.~\ref{sec:Running}. Hence, starting with the low energy value of $\lambda$, known from the Higgs boson mass, 
its value at higher energies is computed in terms of $v_3$ and $v^{\prime}$. The input parameters, in the $\rm \overline{MS}$ scheme, that we use here are~\cite{Huang:2020hdv}
\begin{equation}
\begin{split}
    \lambda(m_Z) &= 0.13947 \pm 0.00045\,,\\
    \alpha_S(m_Z) &= 0.1179 \pm 0.0009\,,\\
    m_t (m_Z) &= (168.26 \pm 0.75) \, \mbox{GeV}.
    \label{eq:CVHPrunning}
\end{split}
\end{equation}
The region of the ($v_3, v^{\prime}$) plane consistent at $2 \sigma$ with $\lambda(v')=0$  is shown in red in Fig.~\ref{fig:Hparity}. That it shuts off for small $v_3$ is understood as follows. Low $v_3$ increases the strong coupling constants at a given scale, thereby enhancing the running of $y_t$, making it decrease faster in the UV than in the SM, hence reducing its impact on $\lambda$, which ends up not crossing zero when $y_t$ runs too fast. One can see that, for the central values in Eq.~\eqref{eq:CVHPrunning}, the condition $\lambda(v^{\prime})=0$ can be satisfied only for $v_3 \gtrsim 10^{10}\, {\rm GeV}$ and $v^{\prime} \gtrsim 6\times 10^{11}\, {\rm GeV}$. Varying the input parameters within their $2\sigma$ uncertainty, $\lambda(v^{\prime})=0$ can be obtained with $v_3$ as low as $\approx 10^7\, {\rm GeV}$. These conditions for Higgs Parity are however incompatible with the results obtained for $e^{\prime}$ as DM candidate, using the central value of $m_u$, as shown by the blue region of Fig.~\ref{fig:Hparity}. (Compare with Eq.~\eqref{eq:DMparam}). As already discussed, a value for $m_u$ larger by $20\%$ would weaken the bounds in Fig.~\ref{fig:resultsDMlargev3} and Fig.~\ref{fig:resultsDMsmallv3}, which is however not sufficient to reconcile Higgs parity and $e'$ DM, as shown by the green region of Fig.~\ref{fig:Hparity}. Higgs Parity in this model would therefore require that measured SM parameters deviate significantly from their central values. Fig.~\ref{fig:Hparity} neglects the threshold corrections to the quartic at $v'$, but it is found to be very small and negative \cite{Dunsky:2019api}, making the situation slightly worse for Higgs parity. Decreasing the theoretical uncertainty on our prediction through the use of two-loop RGEs would be interesting as well, although we do not expect a different conclusion; we leave this for future work.

A non-minimal theory, with additional heavy fermion states coupled to the Higgs, with mass well below $v'$, would allow $\lambda(v') =0$ to be consistent with $e'$ DM from sequential freeze-in.  Thus parity could be spontaneously broken by the radiative Higgs Parity mechanism in non-minimal theories. Froggatt-Nielsen type theories of flavor contain such heavy fermions; if their masses are well below $v'$ the scale of spontaneous flavor symmetry breaking is relatively low.

%%%%%%%%%%%%%%%%%%%%%%%%%%%%%%%%%%%%%%%
\section{Conclusion}\label{sec:Conclusion}

We have shown that certain models based on Parity solutions to the strong CP problem have a DM candidate already embedded in their particle spectrum. Having as a benchmark the model detailed in Ref.~\cite{Bonnefoy:2023afx}, where the full gauge group of the SM is copied. We have discussed the parameters of the model, stressing that Parity leaves little freedom, making the model very predictive. There are two free parameters in addition to the SM ones: the scale at which parity is broken, $v^{\prime}$, which is also the scale of mirror electroweak symmetry breaking, and the scale $v_3$ at which the two color group break to their diagonal subgroup. We computed the impact of broken parity on the RGEs of the model, and we then studied the unavoidable kinetic mixing between the SM $U(1)_Y$ gauge group and its mirror copy, which plays a relevant role for DM direct detection. 

We identified the mirror electron $e'$ as a good DM candidate, while the mirror up quark $u^{\prime}$ can form fractionally charged bound states with SM quarks after QCD confinement, being therefore excluded by several experimental searches. The closeness in mass of $e^{\prime}$ and $u^{\prime}$ strongly determines the DM production mechanism. We find that production of $e^{\prime}$ DM from the SM bath, with a sufficiently suppressed $u'$ abundance, can occur via a sequential freeze-in mechanism through an out-of-equilibrium bath of mirror photons. 
This can occur only in the blue wedge-shaped region in the $(v_3,v')$ plane of Fig.~\ref{fig:Hparity}. Hence, the mass of the mirror electron is in the range $[5-250]$ TeV and the $SU(3) \times SU(3)'$ breaking scale is in the range $[5-500]$ TeV. At any point in this blue wedge region, the reheating temperature must be close to $10^{-7} v'$, and hence is predicted to be low, in the range $[0.1 - 5]$ TeV. We noted that knowing the mass of the up quark with good precision is crucial to make a robust prediction, therefore we commented on the possibility that the precision of lattice determinations is underestimated. 

The blue wedge-shaped region of Fig.~\ref{fig:Hparity} for $e'$ DM is not large and has several observational signals associated with it. Near the vertical edge of the wedge, at $v_3 = 5$ TeV, there are colored states associated with $SU(3) \times SU(3)'$ breaking that may be accessible to future collider experiments, as discussed in \cite{Bonnefoy:2023afx}. Close to the long sloped edge of the wedge, $e'$ DM may be discovered by direct detection, via kinetic mixing of our photon with the mirror photon. Higher in the wedge $v'$ increases and the $u'/e'$ mass ratio falls; the abundance of the fractionally charged hadrons $h$, containing $u'$ increases, leading to signals of this exotic DM component as discussed in \cite{Dunsky:2018mqs}. Finally, throughout the wedge, $e'$ DM is self interacting, with a long-range mirror electromagnetic force that is precisely predicted, and this may lead to future observational signals in large scale structure.

The low reheating temperature requires a late production of the cosmological baryon asymmetry. The theory satisfies two key requirements for leptogenesis: heavy neutral fermions ($\nu'$) with Yukawa couplings to SM leptons, and a reheat temperature above the electroweak phase transition. Generating sufficient baryons at such late times requires $\nu'$ degeneracy, to enhance the asymmetry, and $\nu'$ masses somewhat larger than the reheat temperature.

Finally, we discussed whether the mechanism of Higgs Parity, which provides the spontaneous breaking of exact parity, can be realised in these models and is compatible with the parameters leading to a good DM candidate. We showed that the current central values for $m_t$ and $\alpha_S$ clearly exclude this possibility. Stretching these values within their $2\sigma$ confidence intervals gets one closer to the region of parameter space where $e'$ can be DM, but overlap also requires a large uncertainty in the up quark mass. If parity is broken spontaneously, either SM parameters are far from their central values, the Higgs Parity theory contains couplings of the Higgs to exotic fermions, or the breaking occurs first in some other sector of the theory and appears as soft breaking in the electroweak sector. 

\section*{Acknowledgments}

We thank the members of the Berkeley Center for Theoretical Physics for several discussions, as well as David Dunsky, Keisuke Harigaya, Soubhik Kumar and Keith Olive for discussions on dark matter direct detection, inflation and reheating. This work is supported by the Office of High Energy Physics of the U.S. Department of Energy under contract DE-AC02-05CH11231 and by the NSF grant PHY-2210390. CS acknowledges additional support through the Alexander von Humboldt Foundation.

%\clearpage
\onecolumngrid

\appendix

%%%%%%%%%%%%%%%%%%%%%%%%%%%%%%%%%%%%%%%
\section{Freeze-in of $\gamma^{\prime}$}\label{app:MPhDistr}

As discussed in Sec.~\ref{sec:FZO} and Sec.~\ref{sec:FZI}, a thermal population of mirror photon is not consistent with the mirror electron being DM. If the latter is produced via a freeze-out mechanism, it leads to an overabundance of mirror up-quarks, forbidden by direct searches. If the $e^{\prime}$ is produced via freeze-in, it leads to a yield larger than the observed one for DM. Therefore, we instead considered a frozen-in population of $\gamma'$ that re-scatters into $e'$. 

Computing the yield of $e'$ through this mechanism then requires tracking the distribution $f_{\gamma'}(t,E)$ of mirror photons as they are produced from gluons thorugh the processes of Fig.~\ref{fig:gggammagamma}. Neglecting all terms proportional to $f_{\gamma'}$ at leading order, the Boltzmann equation reads
\beq
\frac{\partial f_{\gamma'}}{\partial t}-HE\frac{\partial f_{\gamma'}}{\partial E}=\frac{1}{E}\int d\Pi_1d\Pi_2d\Pi_3\(2\pi\)^4\delta^{(4)}(\p_1+\p_2-\p_3-\p_{\gamma'})\abs{{\cal M}_{12\to 3\gamma'}}^2f_1(t,p_1)f_2(t,p_2)
\eeq
with $d\Pi_i\equiv\frac{d^3\vec p_i}{(2\pi)^3 2E_i}$ and $p\equiv |\vec p|$. In our case, the particles labeled by $1,2$ are SM gluons in thermal equilibrium, and $3$ is a mirror photon or SM gluon in the final state. Anticipating that the mirror photons will create heavy $e'$, we focus on high energy mirror photons, and hence on high energy gluons. In addition, since the gluons are in thermal equilibrium, we can approximate $f_{1,2}(t,p)\sim e^{-\abs{p}/T}$ in the comoving frame. Then, via momentum conservation, we find
\beq
f_1(t,p_1)f_2(t,p_2)=e^{-(p_3+E)/T} \ .
\eeq
Consequently, we can first compute
\beq
\int d\Pi_1d\Pi_2\(2\pi\)^4\delta^{(4)}(\p_1+\p_2-\p_3-\p_{\gamma'})\abs{{\cal M}_{12\to 3\gamma'}}^2,
\eeq
which is Lorentz invariant. In the center of mass frame, it reads
\beq
\int d\cos\theta\frac{1}{16\pi}\abs{{\cal M}\(s,t=-\frac{s}{2}\(1+\cos\theta\)\)}^2 \ .
\eeq
In the present case, we have computed the cross sections to be,
\beq
\label{relevantGammaPrimeXS}
\bead
&\abs{{\cal M}_{gg\to \gamma'\gamma'}}^2=\frac{e'^4 g_s^4 \left(145586 s^4+516433 s^3 t+725757 s^2 t^2+450254 s t^3+141256 t^4\right)}{114307200 \pi ^4 m_{u'}^8} \\
&\abs{{\cal M}_{gg\to g\gamma'}}^2=\frac{e'^2 g_s^6 \left(145586 s^4+516433 s^3 t+725757 s^2 t^2+450254 s t^3+141256 t^4\right)}{121927680 \pi ^4 m_{u'}^8} \ .
\eead
\eeq
We integrate over $\cos\theta$ and express the result in terms of $s$ in order to have the expression in any Lorentz frame, including the comoving frame where we know the gluon distribution:
\beq
gg\to \gamma'\gamma' \ : \frac{224881 e'^4 g_s^4 s^4}{4572288000 \pi ^5 m_{u'}^8} \ , \qquad
gg\to g\gamma' \ : \quad \frac{224881 e'^2 g_s^6 s^4}{4877107200 \pi ^5 m_{u'}^8} \ . 
\eeq
Performing the last integration over $\vec p_3$, we obtain
\beq
\int d\Pi_3s^4e^{-(p_3+E)/T}=\frac{2E^4e^{-E/T}}{\(2\pi\)^2}\int dp_3d\cos\theta \, p_3^5\(1-\cos\theta\)^4e^{-(p_3+E)/T}=\frac{1536E^4T^6e^{-E/T}}{\(2\pi\)^2} \ ,
\eeq
where we have used the rotational invariance of the system to align $\vec p_{\gamma '}$ with the z-axis. Summing over the two production channels, we find
\beq
\frac{\partial f_{\gamma'}}{\partial t}-HE\frac{\partial f_{\gamma'}}{\partial E}= \frac{224881 e'^2 g_s^4\left(16 e'^2+15 g_s^2\right) T^6 E^3  e^{-\frac{E}{T}}}{190512000 \pi ^7 m_{u'}^8}.
\eeq

We then wish to manipulate this equation into a numerically-friendly form. First, we convert the time derivative into a temperature derivative using $s=\frac{2\pi^2}{45}h_{\rm eff}T^3$,
\beq
\frac{\partial{f_{\gamma'}}}{\partial t} = \frac{\partial{f_{\gamma'}}}{\partial T}\frac{\partial T}{\partial s}\frac{\partial s}{\partial t} =-\bigg(\frac{8\pi^3}{90}\bigg)^{1/2}\frac{T^3}{M_{\rm pl}}\frac{h_{\rm eff}(T)}{g_{*}^{1/2}(T)}\frac{\partial{f_{\gamma'}}}{\partial T}\,,
\eeq
where $g_*^{1/2}\equiv\frac{h_{\rm eff}}{g_{\rm eff}^{1/2}}\(1+\frac{T}{3h_{\rm eff}}\frac{dh_{\rm eff}}{dT}\)$, and we write the Hubble parameter in terms of temperature,
\beq
H(T) = \left(\frac{\pi^3 g_{\rm eff}(T)}{90}\right)^{1/2} \frac{T^2}{M_P}.
\eeq
The equation becomes
\beq
    T\frac{\partial{f_{\gamma'}}}{\partial T} +\alpha_1(T) E\frac{\partial f_{\gamma'}}{\partial E}\\ = -\alpha_2(T) T^4E^3\, e^{-E/T},
\eeq
where we define $\alpha_1(T) = \left(\frac{g_{\rm eff}(T)g_{*}(T)}{8 h_{\rm eff}^2(T)}\right)^{1/2}$ and $\alpha_2(T) = \frac{224881 e'^2 g_s^4\left(16 e'^2+15 g_s^2\right)}{190512000 \pi ^7 m_{u'}^8}\frac{g_{*}^{1/2}(T)}{h_{\rm eff}(T)}\sqrt{\frac{90}{8\pi^3}}M_P$ for simplicity. Because we know that the freeze-in is UV-dominated, we can take $\alpha_1$ and $\alpha_2$ to be constant, giving us the analytical solution,
\beq
\begin{split}
f(T,E) = \frac{\alpha_2 E^3 T^{-3 \alpha_1}}{\alpha_1-1}&\left[T^{3 \alpha_1+4} \left(\frac{E}{T}\right)^{\frac{3 \alpha_1+4}{1-\alpha_1}} \Gamma \left(\frac{3 \alpha_1+4}{\alpha_1-1},\frac{E}{T}\right)\right.\\
&\qquad\left.-T_{\rm RH} ^{3 \alpha_1+4} \left(E\, T_{\rm RH} ^{\alpha_1-1} T^{-\alpha_1}\right)^{\frac{3 \alpha_1 +4}{1-\alpha_1}} \Gamma \left(\frac{3 \alpha_1+4}{\alpha_1-1},T^{-\alpha_1} E\, T_{\rm RH} ^{\alpha_1-1}\right)\right]\,.
\end{split}
\label{eq:DarkPhotonDistr}
\eeq
where $\Gamma$ is the so-called incomplete gamma function. In our Boltzmann codes, we are using the above formula even for energies larger than $m_{u'}$, although Eqs.~\eqref{relevantGammaPrimeXS} only hold for $s\leq m_{u'}^2$. However, as in footnote~\ref{footnoteValidityXS}, energies larger than $m_{u'}$ essentially do not affect the $\gamma'$ distribution and the resulting $e'$ relic abundance.

%%%%%%%%%%%%%%%%%%%%%%%%%%%%%%%%%%%%%%%%%%%%%%%%%%%%%%%%%%%%%%%%%%%%

\twocolumngrid

\cleardoublepage
\bibliographystyle{apsrev4-1_title}
\bibliography{biblio.bib}

\end{document}